Article type: Article

# Isolating hydrogen in hexagonal boron nitride bubbles by a plasma treatment

Li He
School of Optical and Electronic Information, Huazhong University of Science and Technology, Wuhan 430074, China
State Key Laboratory of Functional Materials for Informatics, Shanghai Institute of Microsystem and Information Technology, Chinese Academy of Sciences, 865 Changning Road, Shanghai 200050, P.R. China
CAS Center for Excellence in Superconducting Electronics (CENSE), Shanghai 200050, China
Email: jason_heli@hust.edu.cn

Huishan Wang, Lingxiu Chen, Xiujun Wang, Hong Xie, Chengxin Jiang
State Key Laboratory of Functional Materials for Informatics, Shanghai Institute of Microsystem and Information Technology, Chinese Academy of Sciences, 865 Changning Road, Shanghai 200050, P.R. China
CAS Center for Excellence in Superconducting Electronics (CENSE), Shanghai 200050, China
Graduate University of the Chinese Academy of Sciences, Beijing 100049, P.R. China
Email: hswang2017@mail.sim.ac.cn, chenlx@shanghaitech.edu.cn, xjwang@mail.sim.ac.cn, xiehong@mail.sim.ac.cn, jiangchx@shanghaitech.edu.cn

Chen Li[#],
Department of Lithospheric Research, University of Vienna, Althanstraße 14, 1090 Vienna, Austria
[#]now at Electron microscopy for Materials research (EMAT), University Antwerpen, Groenenborgerlaan 171, 2020 Antwerpen, Belgium
Email: lichen0320@gmail.com

Kenan Elibol,[$] Jannik Meyer[&]
University of Vienna, Physics department, Boltzmanngasse 5, 1090 Vienna, Austria
[$] Now at School of Chemistry, Trinity College Dublin, Dublin 2, Ireland and Advanced Microscopy Laboratory, Centre for Research on Adaptive Nanostructures and Nanodevices, Dublin 2, Ireland
Email: elibolk@tcd.ie
[&] Now at Institute for Applied Physics and Natural and Medical Sciences Institute, University of Tübingen, Tübingen, Germany
Email: jannik.meyer@uni-tuebingen.de

Kenji Watanabe, Takashi Taniguchi
National Institute for Materials Science, 1-1 Namiki, Tsukuba, 305-0044, Japan
Email: WATANABE.Kenji.AML@nims.go.jp, taniguchi.takashi@nims.go.jp

Zhangting Wu, Wenhui Wang, Zhenhua Ni
Department of Physics, Southeast University, Nanjing 211189, P. R. China
Email: 18795895729@163.com, duandaiwang@163.com, nzhfly@163.com

Xiangshui Miao, Chi Zhang, Daoli Zhang*
School of Optical and Electronic Information, Huazhong University of Science and Technology, Wuhan Hubei 430074, China
Engineering Research Centre for Functional Ceramics, the Ministry of Education, Huazhong University of Science and Technology, 1037 Luoyu Road, Hongshan District, Wuhan Hubei 430074, P. R. China
Email: miaoxs@hust.edu.cn, zhangchi_hust@hust.edu.cn, zhang_daoli@hust.edu.cn

Haomin Wang*
State Key Laboratory of Functional Materials for Informatics, Shanghai Institute of Microsystem and Information Technology, Chinese Academy of Sciences, 865 Changning Road, Shanghai 200050, P.R. China
CAS Center for Excellence in Superconducting Electronics (CENSE), Shanghai 200050, China
Email: hmwang@mail.sim.ac.cn

Xiaoming Xie
State Key Laboratory of Functional Materials for Informatics, Shanghai Institute of Microsystem and Information Technology, Chinese Academy of Sciences, 865 Changning Road, Shanghai 200050, P.R. China
University of Chinese Academy of Sciences, Beijing 100049, Peoples Republic of China
CAS Center for Excellence in Superconducting Electronics (CENSE), Shanghai 200050, China
School of Physical Science and Technology, ShanghaiTech University, 319 Yueyang Road, Shanghai 200031, P. R. China
Email: xmxie@mail.sim.ac.cn

* Correspondence and requests for materials should be addressed to D.Z. (email: zhang_daoli@hust.edu.cn) or to H.W. (email: hmwang@mail.sim.ac.cn)

Atomically thin hexagonal boron nitride (*h*-BN) is often regarded as an elastic film that is impermeable to gases. The high stabilities in thermal and chemical properties allow *h*-BN to serve as a gas barrier under extreme conditions. In this work, we demonstrate the isolation of hydrogen in bubbles of *h*-BN via plasma treatment. Detailed characterizations reveal that the substrates do not show chemical change after treatment.

The bubbles are found to withstand thermal treatment in air, even at 800 °C. Scanning transmission electron microscopy investigation shows that the *h*-BN multilayer has a unique aligned porous stacking nature, which is essential for the character of being transparent to atomic hydrogen but impermeable to hydrogen molecules. In addition, we successfully demonstrated the extraction of hydrogen gases from gaseous compounds or mixtures containing hydrogen element. The successful production of hydrogen bubbles on *h*-BN flakes has potential for further application in nano/micro-electromechanical systems and hydrogen storage.

**Introduction**

Hexagonal boron nitride (abbreviated as *h*-BN in the following) is a remarkable two-dimensional mesh consisting of alternating *sp*$^2$-bonded boron and nitrogen atoms.[1] As a wide bandgap insulator,[2] *h*-BN is regarded as an ideal substrate and tunneling barrier for graphene devices because of its atomically smooth surface free of dangling bonds and charge traps.[3] In addition, the exceptional thermal and chemical stabilities of *h*-BN enable it to function as an anti-oxidation layer, even at temperatures as high as 1100 °C.[4] Monolayer *h*-BN sheets remains stable up to 800 °C in air, whereas graphene oxidizes at temperature above 500 °C.[5] This makes *h*-BN a top choice for ultrathin anti-oxidization coatings. In addition to the potential for application in electronics and corrosion-proof coatings, *h*-BN holds promise for selective membrane applications under extreme conditions due to its ultimate thinness, flexibility and mechanical strength.[4, 6-8]

Similar to graphene,[9] *h*-BN is elastic and impermeable to all gases. High-quality *h*-BN sheets have a Young's modulus of ~0.85 TPa and an excellent fracture strength of ~70 GPa.[8] The sheets can withstand stretching of up to 20%.[8] This feature makes *h*-BN a superb material for application in micro-electromechanical systems (MEMSs). A perfect *h*-BN crystal, even when just a monolayer thick, is typically impermeable to most atoms and molecules under ambient conditions. Only accelerated hydrogen atoms possess sufficient kinetic energy to penetrate through *h*-BN planes without damaging the lattices.[10, 11] These properties allow bubbles with various shapes to be produced on *h*-BN.[12] Recently, several reports have described the trapping of gases by graphene.[12-16] By forming bubble structures, the adhesion energy between graphene sheets and substrate can be precisely measured.[17, 18] A voltage-controlled graphene micro-lens with variable focuses[19] and a laser-driven nano-engine[20] have been realized, as the curvature and shape of the bubbles can be controlled to a great extent by variation in the voltage. Although *h*-BN has a similar honeycomb lattice to graphene, little investigation of *h*-BN bubbles for gas isolation has been reported. Compared to graphene, *h*-BN is insulating and more chemically stable, which may greatly extend its application in extreme conditions.

There have been several reports about plasma treatment on 2D materials. Most of them are about plasma applications in surface etching,[21, 22] low-temperature growth [23, 24] and surface hydrogenation.[25] In this work, we demonstrate the isolation of hydrogen in *h*-BN bubbles on the microscale via plasma treatment. The dimension of the bubbles can be regulated by altering the treatment conditions. The diameter of the bubbles ranges from tens of nanometers to several micrometers. The masses of $H_2$ molecules inside the bubbles are believed to have been converted to atomic hydrogen by the plasma, which permeate through the *h*-BN plane and is then captured inside the *h*-BN interlayers. Re-formation of $H_2$ molecules deforms the *h*-BN structure to form bubbles, as the $H_2$ molecules cannot escape from the *h*-BN net. Our demonstration provides an effective method for fabricating bubbles on *h*-BN and may also be an approach for extracting/storing hydrogen in *h*-BN.

**Results**

**Bubble-formation on *h*-BN.**
Fig. 1a shows a schematic illustration of the experimental design for the bombardment of the *h*-BN surface by plasma in different gaseous environments. As shown in Fig. 1, three different gases (Ar, $O_2$ and $H_2$) were fed into the chamber to examine the ability of different elements to penetrate *h*-BN. The flow rate was kept at 3 sccm, while the pressure was kept at ~3 Pa. Plasmas were then generated by a RF generator with a power of 100 W. All *h*-BN flakes were treated with the plasma at 350 ℃ for 150 minutes. Atomic force microscopy (AFM) images of the *h*-BN flakes were taken after the plasma treatment and are given in Fig. 1b-d (the corresponding optical images are given in Supplementary Fig. 1). As shown in Fig. 1b-d, the *h*-BN flakes treated with argon or oxygen plasma appear to be intact, while a large amount of bubbles appeared on the *h*-BN surface that was treated with the hydrogen plasma (H-plasma). A series of experiments including structural characterization by transmission electron microscopy (TEM) (see Supplementary Fig. 2-3), chemical characterization by X-ray photoelectron spectroscopy (XPS) (see Supplementary Fig. 4-5), Energy-dispersive X-ray spectroscopy (EDX) (see Supplementary Fig. 6-7), Fourier Transform infrared spectroscopy (FTIR) (see Supplementary Fig. 8) and mass spectrometry (see Supplementary Fig. 9-10) have been conducted to verify the *h*-BN bubble structure intact and the presence of hydrogen gases inside the bubbles.

To understand the mechanism behind the interesting bubble formation is important. It is well known that gaseous molecules subjected to a strong electromagnetic field decompose into a plasma state, producing masses of charged ions whose particle radii are considerably decreased relative to the original atoms. The hexagonal honeycomb lattice of *h*-BN possesses a lattice constant ($a_{h\text{-BN}}$) of ~250.4 pm.[26] A hydrogen atom has a diameter of ~240 pm (with a van der Waals radius of ~120 pm[27, 28]), which is less than the value of $a_{h\text{-BN}}$ and could pass through the valence electrons concentrated around the BN atoms. Moreover, large amounts of protons, which have a diameter of ~1.68 fm,[29, 30] are generated in the plasma state. These extremely small particles can easily penetrate the BN. The accelerated energetic protons and electrons could then recombine to form hydrogen molecules. These hydrogen molecules are not easily able to pass through the *h*-BN honeycomb due to their large kinetic diameter of ~289 pm,[31] which is substantially larger than $a_{h\text{-BN}}$. Thus, the pressure from the trapped gas causes bubbles to form in the surface of the *h*-BN flakes. In contrast, the atomic diameters of both argon (~376 pm) and oxygen (~304 pm) (with the van der Waals Radii of ~188 pm and ~152 pm, respectively[27, 28]) are considerably larger than $a_{h\text{-BN}}$, and as a result, they cannot penetrate the *h*-BN honeycomb barrier.

The mechanism of unimpeded penetration of atomic hydrogen into the *h*-BN interlayers can also be ascribed to two additional reasons. The first one is the polarization of the electron density distribution in this kind of 2D material. Recent experiments have demonstrated that *h*-BN possesses the best porosity among 2D materials.[11] The concentration of valence electrons around the N atoms results in strong polarization of the BN bonds, making *h*-BN much more porous than graphene and MoS$_2$. The second reason is the stacking order. It has previously been confirmed in the former work that bi- and tri-layered graphene are also non-conductive to protons due to their staggered lattices (AB stacked), in which the electron cloud of the first layer clogs the pores of the second.[11] In contrast, multilayered *h*-BN exhibit excellent proton conductivity probably due to its favoring the AA′ stacking structure, which has been found to be the most stable one among the five possible stacking structures (AA′/AA/AB/AB′/A′B) in recent calculations.[32] The aligned pores in the out-of-plane direction also allow protons to move unimpededly throughout the space. This characteristic structure of multilayer *h*-BN could facilitate its use in in hydrogen storage applications.

To study the stacking order of the *h*-BN multilayer, a scanning transmission electron microscopy (STEM) investigation was carried out. The *h*-BN flakes are firstly exfoliated onto a quartz substrate, and then annealed to remove the residue of resist in an oxygen flow at 800 ℃. After cooling down to room temperature, a group of wrinkles are found on the surfaces. The wrinkles, which are found along the armchair direction of *h*-BN crystals,[33, 34] help us determine the cutting direction of the focused ion beam (FIB). The specimen obtained for the STEM investigation is similar to that shown in Supplementary Fig. 3a.

Fig. 2a illustrates that the incident electron beam (e-beam) for the STEM investigation is parallel with the [11$\bar{2}$0] crystallographic axis of *h*-BN. The cross-sectional schematic of the armchair edge-on view of *h*-BN multilayer is presented in Fig. 2b. Fig. 2c is a Wiener filtered STEM medium angel annular dark-field (MAADF) image showing the edge-on view configuration of *h*-BN multilayer, which reveals the layer stacking of the *h*-BN. The intensity profiles along two selected lines (red/blue) in Fig. 2c are showed in Fig. 2d, exhibiting apparent synchronism of the peak positions. Besides, the average peak-to-peak distances measured in red and blue profiles are both ~2.26 Å, which are close to 2.166 Å expected in *h*-BN. The minor deviation of the spacing in STEM measurement might due to the limitation of the scale calibration of the instrument. This result shows that the stacking order of our *h*-BN is either AA′ or AA in our *h*-BN multilayer, which is in line with earlier predictions. Beside the cross-sectional STEM image, we also carried out a plan-view STEM investigation (shown in Supplementary Fig. 18) of another $h$-BN multilayer. The image confirms that the $h$-BN multilayer we fabricated is in AA′ stacking. These images provide direct evidence that the multilayered planes in $h$-BN are not staggered so that the electron cloud of each layer does not block the pores of the successive layer. As such, atomic hydrogen could tunnel through multilayered $h$-BN.

**Control on size of $h$-BN bubbles.**
Both the duration and the environmental temperature of plasma treatment have prominent effects on the density and sizes of bubbles, which provide a feasible way to fabricate $h$-BN bubbles with controllable dimensions. To understand the dependence of the bubble dimensions on the duration of the H-plasma treatment, we varied the treatment duration while holding the other conditions constant (100 W RF, 350 ℃, 3 sccm $H_2$, ~3 Pa) and examined the variations in the dimensions of the generated bubbles. The results of this experiment are given in Fig. 3. As shown in Fig. 3a, the $h$-BN samples obtained in different duration of H-plasma treatment possessed bubbles with obviously different densities and size distributions. The bubbles treated for 90 min were generally smaller than 100 nm in diameter and had a higher density in the distribution, while plasma treatment for 150 minutes led to a lower bubble density and a larger bubble size with diameters greater than 3 μm. The dependence of diameter distribution on treatment duration is plotted in Fig. 3b. Extended plasma treatment favors the formation of larger bubbles. An H-plasma treatment for 500 min could fabricate larger bubbles. As shown in Supplementary Fig. 11a-c, bubbles with diameters of ~20 μm were obtained on $h$-BN.

In addition to the treatment duration, the environmental/sample temperature is another key factor that influences the dimensions of the bubbles. Thus, the sample temperature was also adjusted while keeping the other conditions constant (100 W RF, 120 minutes, 3 sccm $H_2$, ~3 Pa) to investigate the variations in the bubble dimensions. Fig. 3c presents the density and diameter fractions of the $h$-BN bubbles for different sample temperatures. Almost all of the bubbles produced at room temperature (30 ℃) were <100 nm in diameter, and the relative fraction of large bubbles gradually increased when the sample temperature was elevated from 30 ℃ to 350 ℃. Meanwhile, the density of the bubbles dramatically decreased with increasing sample temperature. Furthermore, the bubble shown at the bottom-left of Supplementary Fig. 12d formed by merging three individual bubbles. The observation reveals that the bubbles can migrate and merge at higher sample temperature. Similar phenomena were also observed in Fig. 3a and Supplementary Fig. 13. The hydrogen molecules inside smaller bubbles, normally possessing higher inner pressure than relatively large bubbles, move more easily at higher sample temperatures in-between $h$-BN interlayers.[15] Hydrogen gas in small bubbles is more likely to pass through the inter-planar channels and combine to form a larger bubble with a decreased inner pressure. Such combination leads to a dramatic decrease in bubble density at high sample temperatures.

We then further explored the relationship between the height/aspect ratio of the bubbles and their diameter. The maximum height, $h_{max}$, and the quotient of $h_{max}$ to the radius $R$ (called the aspect ratio) of each bubble are taken from all observed $h$-BN bubbles by AFM. The results are presented in Fig. 3d. It is found that $h_{max}$ approximately increases linearly with bubble diameter, while the aspect ratio remains almost unchanged.

**Mechanisms of the plasma driven effect.**
In order to understand the mechanism behind the plasma process, exploring hydrogen plasma itself is essential.[35] We carried out modeling of hydrogen plasma in a tube using COMSOL Multiphysics. The spatial distribution of different parameters in the tube are plotted in Supplementary Fig. 14. As shown in Supplementary Fig. 14e-f, k-l, the ion density at the sample area decreases while the thermal velocity increases, when the sample was heated from room temperature to 350 °C. The results indicate that protons in H-plasma can be injected much faster into h-BN interlayers at a higher environmental temperature with comparatively lower spatial density. This simulation exhibits the variation tendency of dimension/density when elevating the environmental temperature (heated by the furnace) of the sample (as shown in Fig. 3c). The furnace temperature (Supplementary Fig. 14a-b), electron density (Supplementary Fig. 14c-d), electron temperature (Supplementary Fig. 14g-h) and pressure (Supplementary Fig. 14i-j) are also given to help in giving images shown plasma-driven kinetics.

It is noticed that the electron density (Supplementary Fig. 14c-d) is 2 orders in magnitude lower than the ion density (Supplementary Fig. 14e-f) at the sample position according to simulation. This phenomenon indicates that the quasi-neutrality of plasma at the sample position is broken. According to the set-up showed in Supplementary Fig. 19a-d, $h$-BN sample was placed 30 cm away from the RF coil (the core region of the plasma). Along the axial direction of tube furnace, there is a steep electron gradient which leads to the unbalance between electron density and ion density, and thus, the quasi-neutrality of plasma is not sustained. To understand this phenomenon further, we then carried out modeling in Debye length when the furnace was set at 350 °C in Supplementary Fig. 14q. Normally, Debye length depends on density of ion number and environmental temperature. Debye length increases when ion number density decreases or when the environmental temperature increases. It is also found that the Debye length, which is much higher than the diameter of the tube (0.042 m), reached its maximum of ~0.5 m at the sample position. It indicates the destruction of quasi-neutrality at the sample position. This also interprets why the electron density is 2 orders lower than the ion density at the sample position.

Besides the sample temperature and treatment duration, the position where the $h$-BN sample was placed also brings different size/distribution of bubble on $h$-BN. The schematic of experimental set-up as well as the results of bubble distribution are exhibited in Supplementary Fig. 15 and Supplementary Fig. 16, respectively. It is obvious that bubbles in the plasma core area are small but dense. while bubbles at other area show relatively large but thin. The electric potential difference (Supplementary Fig. 14m-n) and the ion number density distribution (Supplementary Fig. 14o-p) are also simulated by COMSOL Multiphysics to analyze the bubble distributions in the supporting information. Obviously, the formation of bubbles is a plasma driven effect. The size and distribution of bubbles on $h$-BN are generally determined by the comprehensive influence of ion density, Debye length and potential gradient at the position where the $h$-BN samples are placed.

**Low temperature AFM measurement.**
To determine the type of gases inside the $h$-BN bubbles, low temperature AFM measurements were carried out to explore their temperature evolution. The experiment was inspired by a very recent literature about bubble structure on bulk transition metal dichalcogenides.[36] Details about measurement are given in the Method section. Fig. 4a shows an optical image of bubbles on a $h$-BN flake. After transferring it into a vacuum chamber equipped with an AFM, the sample was cooled down gradually. Fig. 4b shows topographic AFM images of the same area measured at 34K and 33K respectively. It is clearly shown that the bubbles deflate at 33K while they are inflated with gases at 34K. As shown in Fig. 4c, the height profiles of line-scans across a bubble (indicated by dashed lines in Fig. 4b) clearly show that the bubble remains at ~34K and disappears at ~33K. The deflating/swelling processes were reproducible via cooling/heating the sample. The highest temperature at which the bubbles become flat was recognized as the transition temperature ($T_{transition}$). We measured a total of 58 bubbles. Three other typical samples are given in Supplementary Fig. 17A-C. The transition temperature of each bubble was recorded when they become flat during cooling. The number of bubbles which have the same $T_{transition}$ is counted. The histogram of $T_{transition}$ as function of temperature at an interval of 1K is plotted in Fig. 4d. As shown in Fig. 4d, $T_{transition}$ of all the bubbles measured can be described as a Gaussian distribution, and exhibits an average value equal to 33.2±3.9 K. This value is very close to $T_{transition}$ of $H_2$ (33.18 K) among all possible gases (see Supplementary Table 2). The result strongly indicates that the gases inside the bubbles are indeed hydrogen molecules.

**Determining the penetration depth of atomic hydrogen in the h-BN multilayer.**
Increasing the power of the RF source may lead to bursting of the bubbles, which leaves many pits on the $h$-BN surface. These pits make it possible to study the penetration depth of atomic hydrogen. Fig. 5a shows an optical micrograph of an $h$-BN flake treated for 180 minutes at 350 °C with an RF power of 400 W. Numerous pits are observed on the $h$-BN surface. Fig. 5b-c depicts the AFM image of a deep pit highlighted by the red box in Fig. 5a. The AFM height image in Fig. 5b provides depth information about the pit, whose depth is given in the insert. The amplitude error image provided in Fig. 5c shows a clearer picture of the morphology around the pit. The bottom of the pit is smooth, which indicates that no atomic hydrogen penetrated the bottom layer. Furthermore, folds are observed in the $h$-BN flaps around the outside of the pit, as illustrated in Fig. 5d. The inset in Fig. 5b shows that the rupture depth of the folded border step is approximately 15.22 nm, which indicates that ~23 layers of $h$-BN were penetrated by atomic hydrogen in the 400 W treatment. Fig. 5e presents the statistics of the penetration depth of atomic hydrogen in the $h$-BN sample shown in Fig. 5a. The penetration depth of the $h$-BN flake ranges from ~11 to ~23 layers. The histogram shown in Fig. 5e reveals that the 400 W RF power provided the H-plasma with enough kinetic energy to penetrate at most ~23 atomic layers underneath the $h$-BN surface.

**Characterizations of Raman spectroscopy.**

Raman spectroscopy was also performed to characterize the $h$-BN bubble. As shown in Fig. 6a, a typical bubble with a round shape was measured. The bubble had a diameter of ~2.36 μm and a height of ~122 nm. The Raman spectrum was measured at three spots from the flat area next to the bubble to the center of the bubble. The results are shown in Fig. 6b. The Raman spectra of $h$-BN clearly show only one peak near 1366 cm$^{-1}$, which corresponds to the $E_{2g}$ vibration mode of $h$-BN. As shown in the inset of Fig. 6b, the $E_{2g}$ peak position and full width at half maximum (FWHM) of each spectrum were extracted and plotted as a function of the measurement site. The peak position in the Raman spectrum appeared at the center of the bubble redshifted (~3 cm$^{-1}$) relative to that of the position outside of the bubble, with almost no change in the FWHM. The observed blue shift in the peak position is due to the presence of stretching strain on the bubble caused by expansion.[37, 38]

**Hydrogen extraction.**

The demonstrated fabrication of hydrogen bubbles on $h$-BN via plasma treatment inspired us to consider the possibility of extracting hydrogen from hydrocarbons and mixtures of gases. Three different gases ($C_2H_2$, $CH_4$ and Ar+$H_2$, 5% of $H_2$) were fed into the chamber for plasma treatment under the same experimental conditions (350 ℃, 100 W, 120 min, ~3 sccm flow rate, ~3 Pa). Excitingly, all three samples formed bubble structures on the $h$-BN surface after the plasma treatment. The experimental illustration and AFM images of the $h$-BN flakes are given in Fig. 7a-c and 7d-f. Considering that the diameters of carbon and argon atoms are much larger than the lattice constant of $h$-BN, it is likely that only atomic hydrogen could penetrate the $h$-BN layers. The results indicate that hydrogen could be separated from hydrocarbons and from the mixture of argon and hydrogen.

As shown in Fig. 7d-f, the bubbles formed after the ethyne plasma treatment had diameters of predominately <300 nm and a high density, while the bubbles produced by the methane plasma possessed a much larger diameter range from 2-4 μm with a much lower density. The exact causes for the phenomena need comprehensive investigation in future. At this moment, the main reason, we believe, could be that the dissociation energy of methane is actually lower than that of the ethyne.[39] When all experiments were carried out under similar conditions, more atomic hydrogen in $CH_4$ plasma can be generated and then more efficiently get into $h$-BN bubbles than the case in ethyne plasma. The plasma treatment with 5% hydrogen in argon also led to the formation of bubbles on the $h$-BN surface but with an extremely low density. To characterize the bubbles further, the plots of $h_{max}$ and the aspect ratio with respect to the bubble diameter are presented in Fig. 7g-i. Similar to the bubbles fabricated by the plasma of pure hydrogen, in all three plots, $h_{max}$ had a nearly linear increasing relationship with the bubble diameter, while the aspect ratio remained nearly unchanged and thus was independent of the bubble diameter. Interestingly, although the bubbles produced from $CH_4$ had much larger diameters (2-4 μm) than those produced from $C_2H_2$ (<300 nm), they had almost the same $h_{max}$ (<10 nm), resulting in an extremely low aspect ratio ($h_{max}/R = $ ~0.005) for the bubbles produced from methane.

**Discussion**

We demonstrate an approach for isolating hydrogen in $h$-BN bubbles by plasma treatment. The dimensions of the bubbles range from nanometers to micrometers and depend on the duration and sample temperature of the plasma treatment. Most of the bubbles undergo minimal changes over 40 weeks. Additionally, the penetration depth of atomic hydrogen reaches approximately 23 $h$-BN layers. The Raman spectra of a bubble reveal that stretching occurs on the bubble surface. Finally, we demonstrate that hydrogen gas can be extracted from hydrocarbons and mixed gases by the plasma treatment and isolated between $h$-BN layers in bubbles. The demonstrated ability to fabricate hydrogen bubbles on $h$-BN flakes might be used to extract hydrogen from hydrocarbon/gas mixtures or to fabricate nano/micro-electromechanical systems.

**Methods**

**Fabrication and characterization of hydrogen bubbles on $h$-BN.**
The experimental setup is given in Supplementary Fig. 19. First, $h$-BN flakes were produced by micromechanical exfoliation of single-crystal $h$-BN and then deposited on quartz substrates that were previously cleaned with oxygen plasma. The substrates with $h$-BN flakes were loaded into a furnace equipped with an RF generator (13.56 MHz, MTI Corporation) with a tunable power from 100 to 400 W. The sample temperature in the sample chamber could be controlled from room temperature to 1000 °C. An advanced pump (GX100N Dry Pumping System, Edwards) was connected to the chamber to control the flow rate. After the samples were heated to the preset temperature, hydrogen (oxygen or argon) gas with a flow rate of 3 sccm (~3 Pa) was introduced into the chamber, and then the plasma was formed. The plasma treatment normally took 90, 120 or 150 min. The $h$-BN flakes were finally removed from the sample chamber for characterization under an optical microscope (Eclipse LV150, Nikon), an atomic force microscope (Dimension Icon, Bruker) and a Raman spectrometer (532 nm, WITec).

**Low temperature AFM measurement of the bubbles.**
After bubbles were produced on $h$-BN flakes, optical microscope and AFM were used to locate their positions. Subsequently, the substrate with the $h$-BN flakes was transferred into the vacuum chamber of a variable temperature SPM (Scanning probe microscope) system (attoDRY 1100 by Attocube). The chamber was filled with $He^4$ in a pressure of 5 mbar for heat exchange. The sample temperature can be controlled in the range from 4 to 300K. All the atomic force microscope (AFM) scanning was conducted in contact mode. In order to minimize the damage to the bubbles, AFM tips with a force constant of 0.02-0.77 N/m were used. The sample was cooled at a rate of 0.5 Kelvin/hour while AFM scanning was carried on the *h*-BN bubbles. The highest temperature at which the bubbles became flat was recorded as the transition temperature. Afterwards, the samples were warmed up for several Kelvins to observe the expansion of bubbles in order to exclude the possible case of gas leakage.

**Scanning transmission electron microscopy.**

Electron microscopy experiments were conducted using a Nion UltraSTEM100 scanning transmission electron microscope, operated at 60 kV in near-ultrahigh vacuum ($2 \times 10^{-7}$ Pa). The beam current during the experiments varied between 8 and 80 pA, corresponding to dose rates of ~5–50 $\times 10^7$ $e^-\text{Å}^{-2}\text{s}^{-1}$. The beam convergence semi-angle was 35 mrad and the semi-angular range of the medium-angle annular-dark-field detector was 60–200 mrad.

**Data availability**

The data that support the finding of this study are available from the corresponding author on request.

**Acknowledgments**

The work was partially supported by the National Key R&D program (Grant No. 2017YFF0206106), the Strategic Priority Research Program of Chinese Academy of Sciences (Grant No. XDB30000000), the National Science Foundation of China (Grant No. 51772317, 51302096), the Science and Technology Commission of Shanghai Municipality (Grant No. 16ZR1442700) the Hubei Provincial Natural Science Foundation of China (Grant No. ZRMS2017000370), and the Fundamental Research Funds of Wuhan City (No. 2016060101010075). K.W. and T.T. acknowledge support from the Elemental Strategy Initiative conducted by the MEXT, Japan and JSPS KAKENHI Grant Numbers JP15K21722. C.L. acknowledges support from the European Union's Horizon 2020 research and innovation program under the Marie Skłodowska-Curie grants No. 656378 – Interfacial Reactions. L.H. acknowledges financial support from the program of China Scholarships Council (No. 201706160037). H.W. and D.Z. thank Y. Gu, Y. Ma, X. Chen (Shanghai Institute of Technical Physics, Chinese Academy of Sciences) for FTIR spectra measurement. L.C. and L.H. thank Q. Liu and Z. Liu (Shanghai Institute of Microsystem and Information Technology, Chinese Academy of Sciences) for measurement in XPS spectra and mass spectra.

**Author Contributions**

H.W. conceived and designed the research. L.H. fabricated the samples and performed the measurements. L.C., C.Z., and H.S.W. assisted with the transfer technique and AFM measurements. C.J. carried out low temperature AFM measurements. X.W. and H.X. provided assistance with substrate fabrication. Z.W., W.W. and Z.N. performed the Raman measurements. C.L., K.E., J.M. carried out STEM measurements. K.W., T.T. fabricated the *h*-BN crystals. L.H., C.L. X.X., X.M., D.Z. and H.W. prepared the manuscript. All authors discussed the results and commented on the manuscript.

**Competing interests**

The authors declare no competing financial interests.

**References**

1. Corso M, Auwarter W, Muntwiler M, Tamai A, Greber T, Osterwalder J. Boron nitride nanomesh. *Science.* **303**, 217-220 (2004).

2. Kubota Y, Watanabe K, Tsuda O, Taniguchi T. Deep ultraviolet light-emitting hexagonal boron nitride synthesized at atmospheric pressure. *Science.* **317**, 932-934 (2007).

3. Dean CR, Young AF, Meric I, Lee C, Wang L, Sorgenfrei S, *et al.* Boron nitride substrates for high-quality graphene electronics. *Nat Nanotechnol.* **5**, 722-726 (2010).

4. Liu Z, Gong Y, Zhou W, Ma L, Yu J, Idrobo JC, *et al.* Ultrathin high-temperature oxidation-resistant coatings of hexagonal boron nitride. *Nat Commun.* **4**, 2541 (2013).

5. Liu L, Ryu S, Tomasik MR, Stolyarova E, Jung N, Hybertsen MS, *et al.* Graphene oxidation: Thickness-dependent etching and strong chemical doping. *Nano Lett.* **8**, 1965-1970 (2008).

**Figures:**

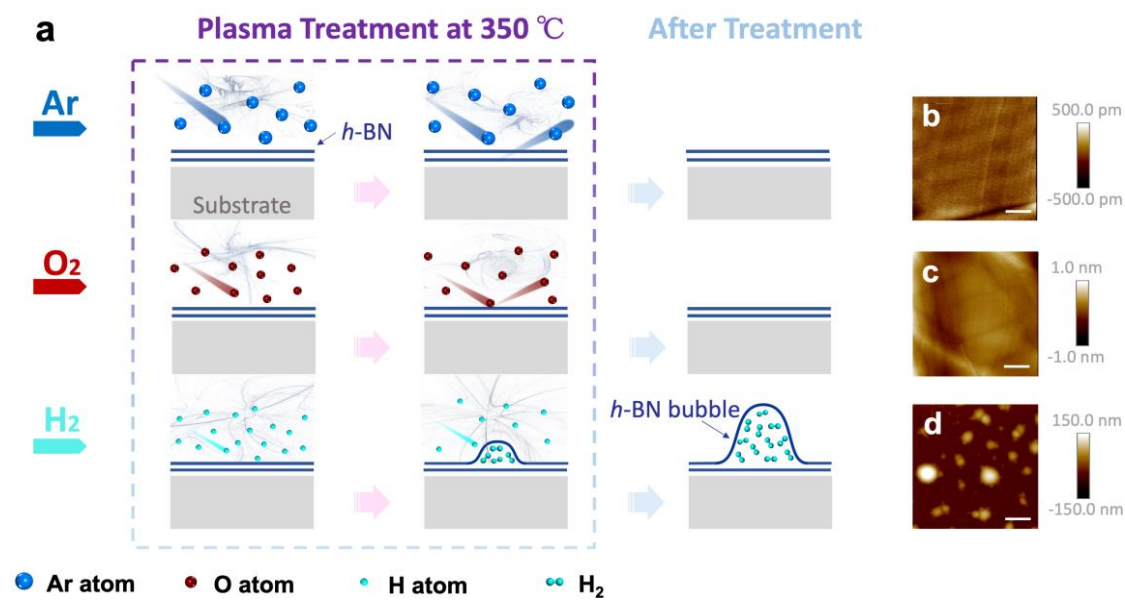

**Figure 1 | Production of bubbles on *h*-BN flakes via plasma treatment. a**, Schematic depicting the plasma treatment of *h*-BN flakes in different atmospheres. All AFM height images of the *h*-BN flakes were taken after plasma treatment in an atmosphere of **b**, argon, **c**, oxygen or **d**, hydrogen. The *h*-BN flakes appear to remain intact after the argon and oxygen plasma treatments, while obvious bubbles are observed on the *h*-BN surface after treatment with the H-plasma. All samples were obtained under similar conditions. Scale bars in **(b-d)** are 4 μm.

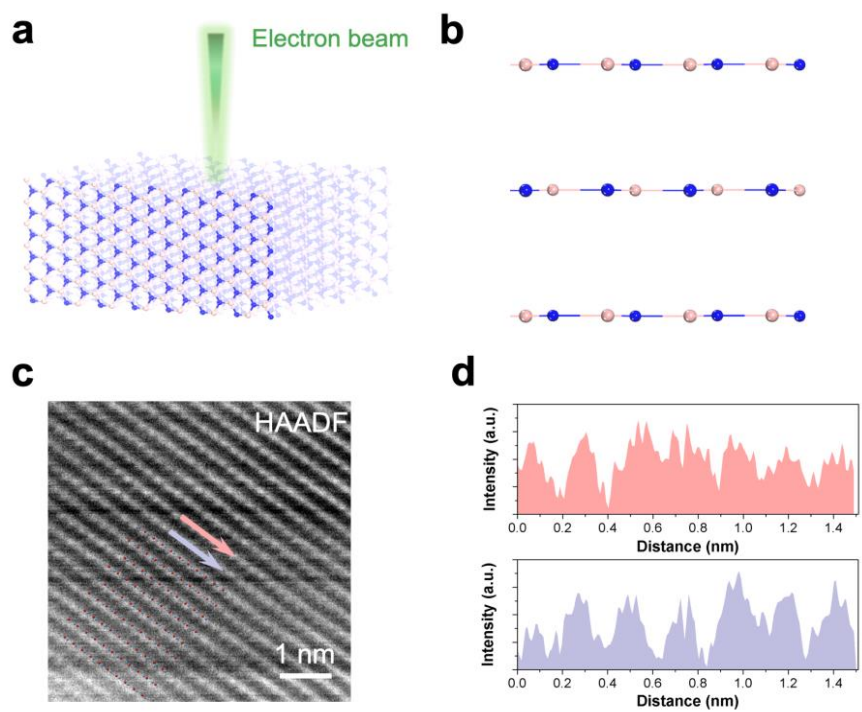

**Figure 2 | STEM image showing the edge-on view configuration of a *h*-BN multilayer. a,** Schematic of a STEM investigation of the *h*-BN multilayer, the incident electron beam was parallel with its $[11\bar{2}0]$ crystallographic axis (the zigzag edge of the *h*-BN layer). **b,** Cross-sectional schematic of the armchair edge-on view of *h*-BN multilayer. The red atoms represent boron atoms while the blue ones represent nitrogen atoms. **c,** STEM-MAADF image, post-processed by a Wiener Filter, showing the edge-on view configuration of a *h*-BN multilayer. **d,** Profiles of image intensity along the red and blue arrows marked in **(c)**, showing the stacking sequence in *h*-BN. Their average peak-to-peak distances are ~2.26 Å, which are in agreement with the lattice spacing of *h*-BN. The synchronism of the peak intensity along the arrows demonstrates the stacking structure of *h*-BN must be either AA′ or AA.

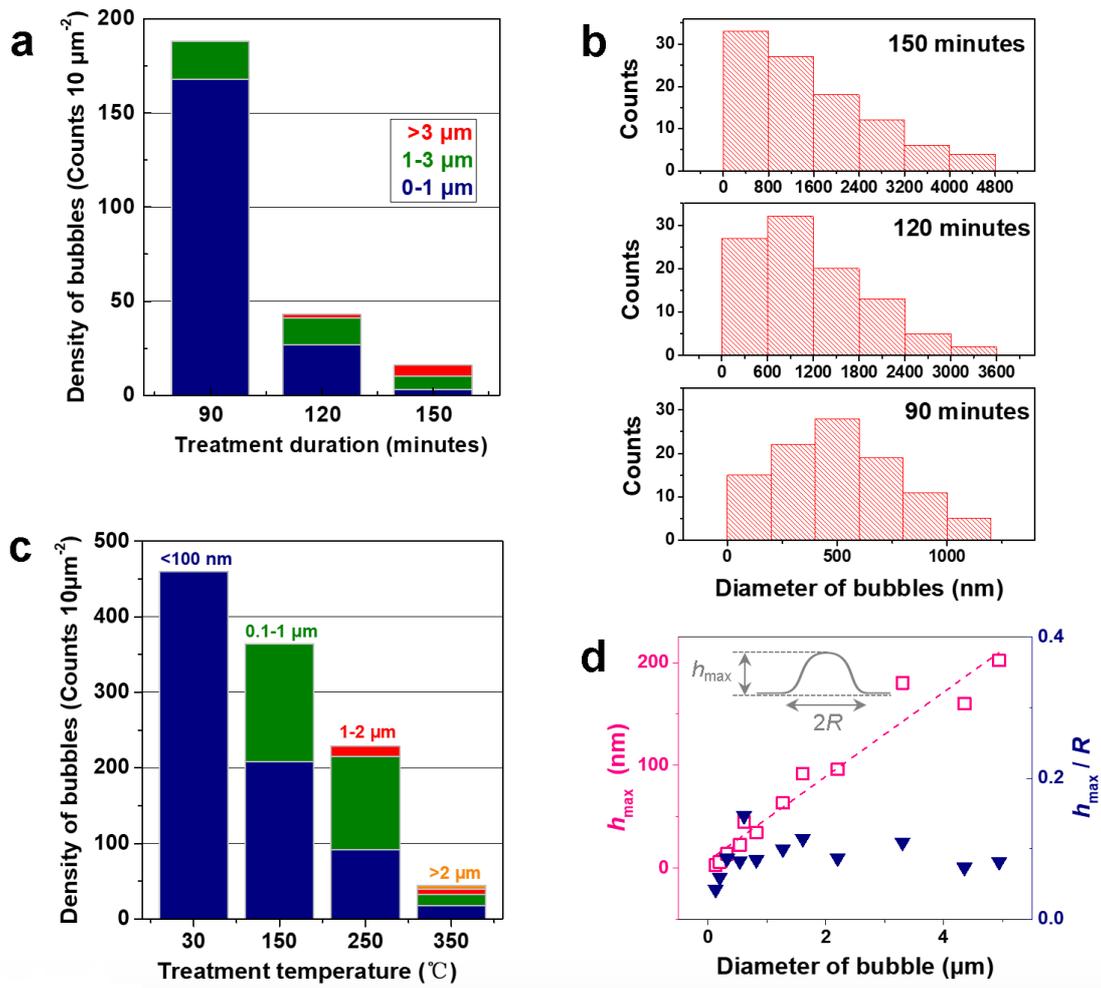

**Figure 3 | Influence of the plasma treatment duration and sample temperature on the dimensions of the bubbles. a**, Density and relative fractions of *h*-BN bubbles within three diameter ranges (0-1 μm, 1-3 μm, and >3 μm) produced by different H-plasma treatment durations; all samples were treated under the typical experimental conditions (100 W RF, 350 °C, 3 sccm $H_2$, ~3 Pa). **b**, Dimension distributions of the *h*-BN bubbles produced after H-plasma treatment for 90, 120, and 150 minutes. **c**, Density and relative fractions of *h*-BN bubbles within four diameter ranges (<100 nm, 0.1-1 μm, 1-2 μm, and >2 μm) produced from treatment at different sample temperatures (30 °C, 150 °C, 250 °C and 350 °C); the other conditions were not changed (100 W RF, 120 minutes, 3 sccm $H_2$, ~3 Pa). **d**, $h_{max}$ of the bubbles produced by the H-plasma and the aspect ratio of the bubbles with respect to the bubble diameter. The inset shows an illustration of $h_{max}$ and *R*.

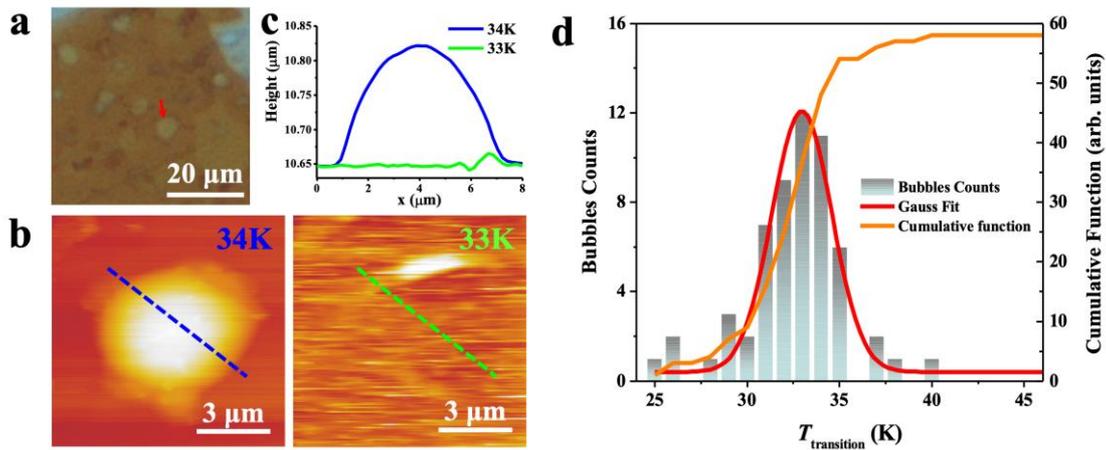

**Figure 4. Swelling and deflating processes of the *h*-BN bubbles containing hydrogen. a,** An optical image of bubbles on a *h*-BN flake, taken under ambient condition, scale bar: 5μm; **b,** Topographic AFM image of a bubble pointed-out by an arrow in **(a)** was measured at 34K and 33K respectively; **c,** The height profiles of line-scan at the same place (indicated by dashed lines in **(b)**) where the bubble remains at ~34K and disappears at ~33K; **d,** Histogram of the transition temperature ($T_{transition}$) at which bubbles collapse. The red line is a Gaussian fit to the data. The yellow line is the histogram cumulative function (right axis).

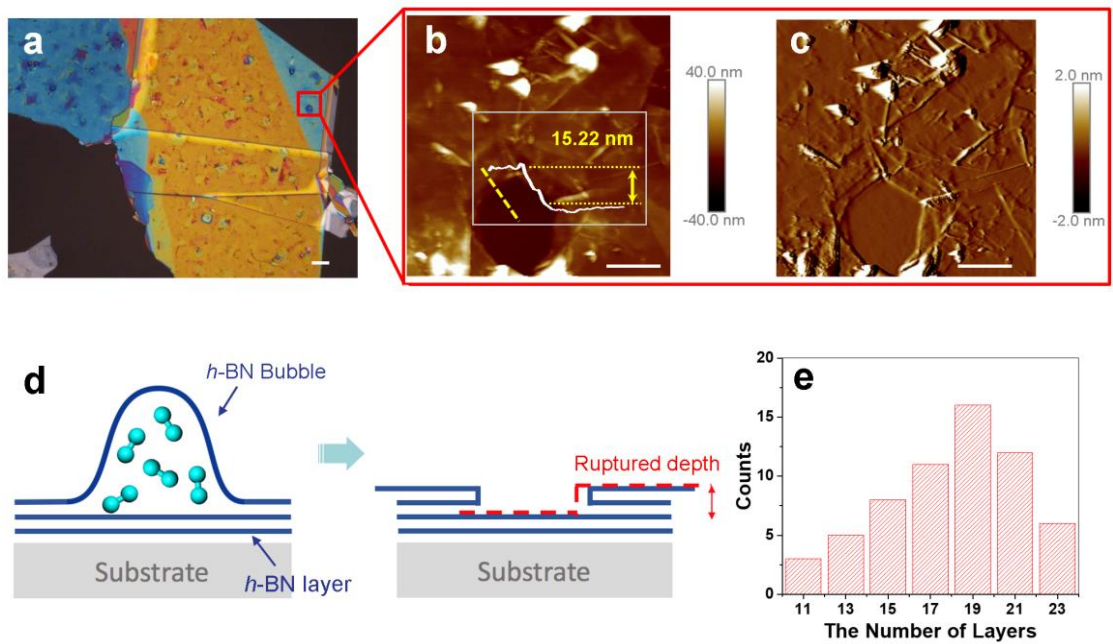

**Figure 5 | Penetration depth of atomic hydrogen into *h*-BN multilayers**. **a**, Optical image of an *h*-BN flake on which the bubbles generated by a 400 W H-plasma treatment ruptured. *h*-BN flaps peeled off the adhered macroscopic films, and the multilayer *h*-

BN sheets tore and then folded. **b**, AFM height and **c**, amplitude error images taken of the red area shown in **(a)**. The bottom of the pit is atomically flat, and no bubble filled with hydrogen molecules was found. The insert in **(b)** is the AFM depth profile along the yellow dashed line. The deviation in the height is approximately 15.2 nm, indicating that ~22 atomic layers were ruptured and folded over the top layer of *h*-BN. The scale bars in **(a-c)** represent 1 μm. **d**, Illustration of the ruptured depth for the folding steps depicted in **(b, c)** caused by the bursting of the *h*-BN bubble. **e**, Depth distribution of penetration into the surface of the *h*-BN flake shown in **(a)**.

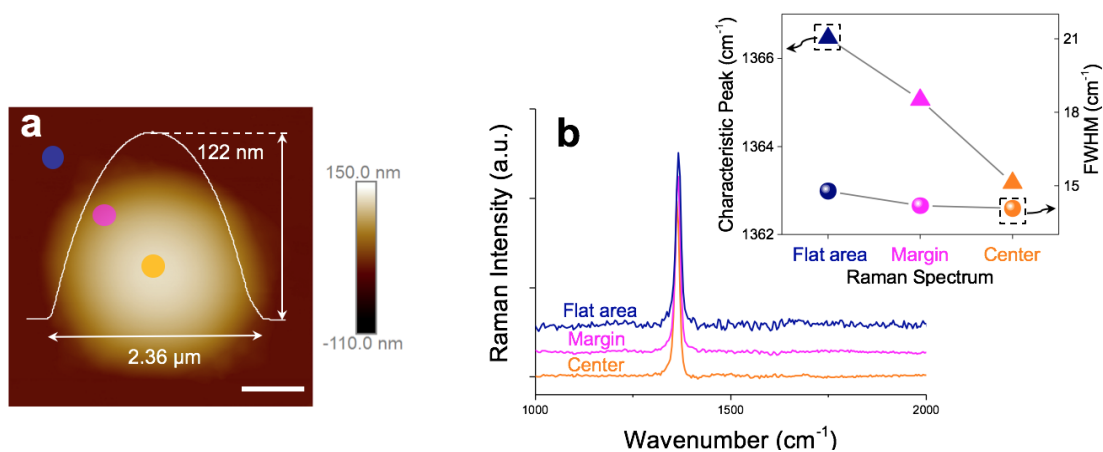

**Figure 6 | Raman spectra taken at different positions of a *h*-BN bubble. a,** AFM image of a typical *h*-BN bubble; the white curve shows the profile across the center of bubble and gives information about its height and diameter. Scale bar, 600 nm. **b,** Raman spectra taken at the positions indicated in the AFM image. The inset shows the variation in the $E_{2g}$ peak position and FWHM between positions. A redshift in the $E_{2g}$ peak position from the flat area to the center of the bubble is observed, while the corresponding FWHM is nearly unchanged.

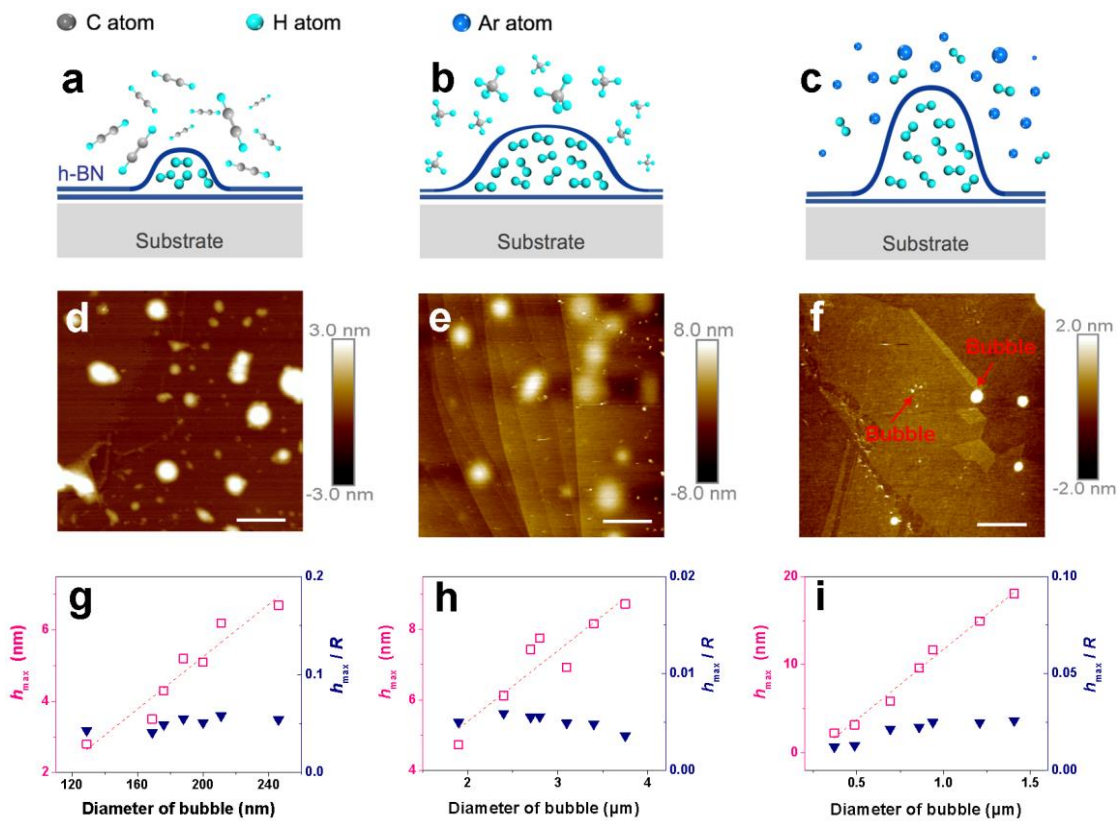

**Figure 7 | Extracting hydrogen from hydrocarbon and mixture gases. a-c**, Schematic illustration of the separation of atomic hydrogen from ethyne ($C_2H_2$), methane ($CH_4$) and a mixture of hydrogen and argon ($Ar+H_2$, 5% $H_2$) by an *h*-BN membrane via plasma treatment. **d-f**, AFM height images of the *h*-BN surfaces treated by a typical plasma process (350 °C, 100 W, 120 min, ~3 sccm, ~3 Pa) in an environment of ethyne, methane and a mixture of hydrogen and argon. Scale bars: **(d)** 600 nm, **(e, f)** 4 μm. **g-h**, $h_{max}$ of the bubbles fabricated via the plasma treatment in $C_2H_2$, $CH_4$ and $Ar+H_2$ (5% of $H_2$), respectively, and their aspect ratios as a function of the bubble diameter.

Supplementary Information for:

# Isolating hydrogen in hexagonal boron nitride bubbles by a plasma treatment

Li He, et al.

# Supplementary Information

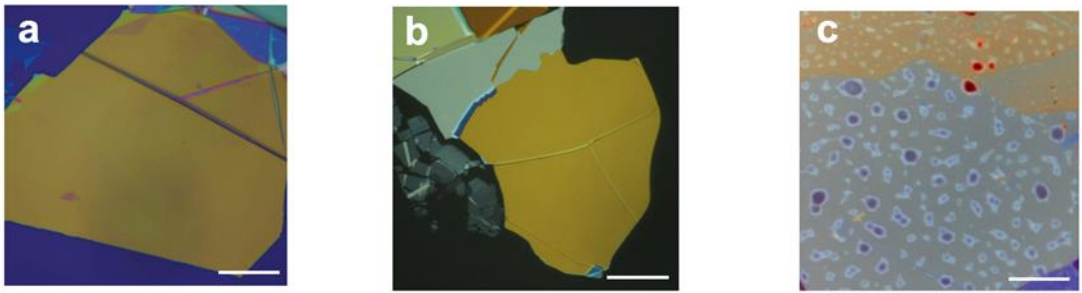

**Supplementary Figure 1 | a-c**, **Optical images of *h*-BN flakes corresponding to the AFM height images shown Fig. 1(b-d), respectively.** Scale bars: 20 μm.

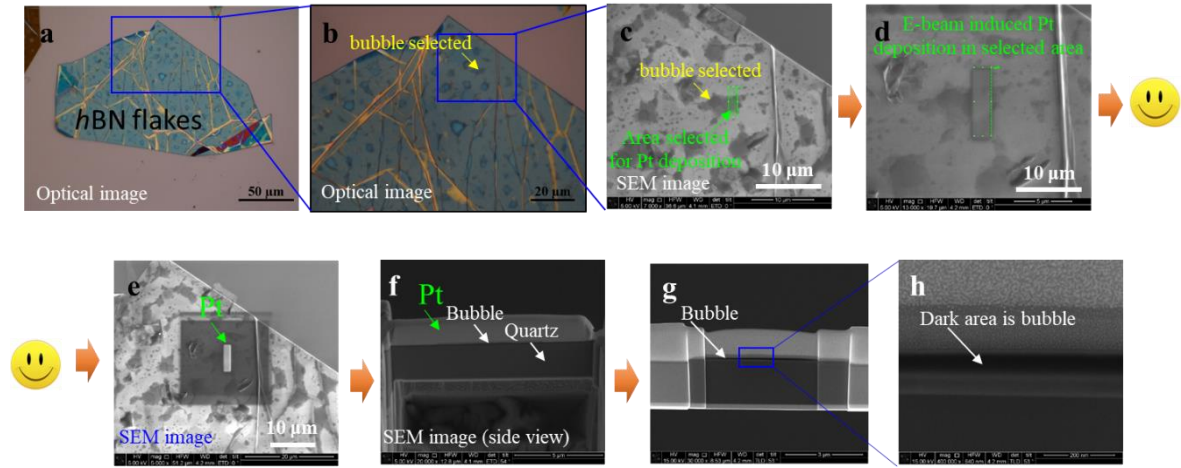

**Supplementary Figure 2 | OM and SEM images acquired during TEM specimen preparation. a-b,** Optical images of *h*-BN flakes with bubbles. The SEM image sequence **(c-h)** corresponds to the chronological steps required for the fabrication of a *h*-BN bubble TEM specimen. The specimen was prepared by dual beam system (Helios NanoLab 600). **c-e,** the SEM images of the top view of the *h*-BN flake with bubbles. **f-h,** Cross-sectional SEM image of the specimen after FIB milling.

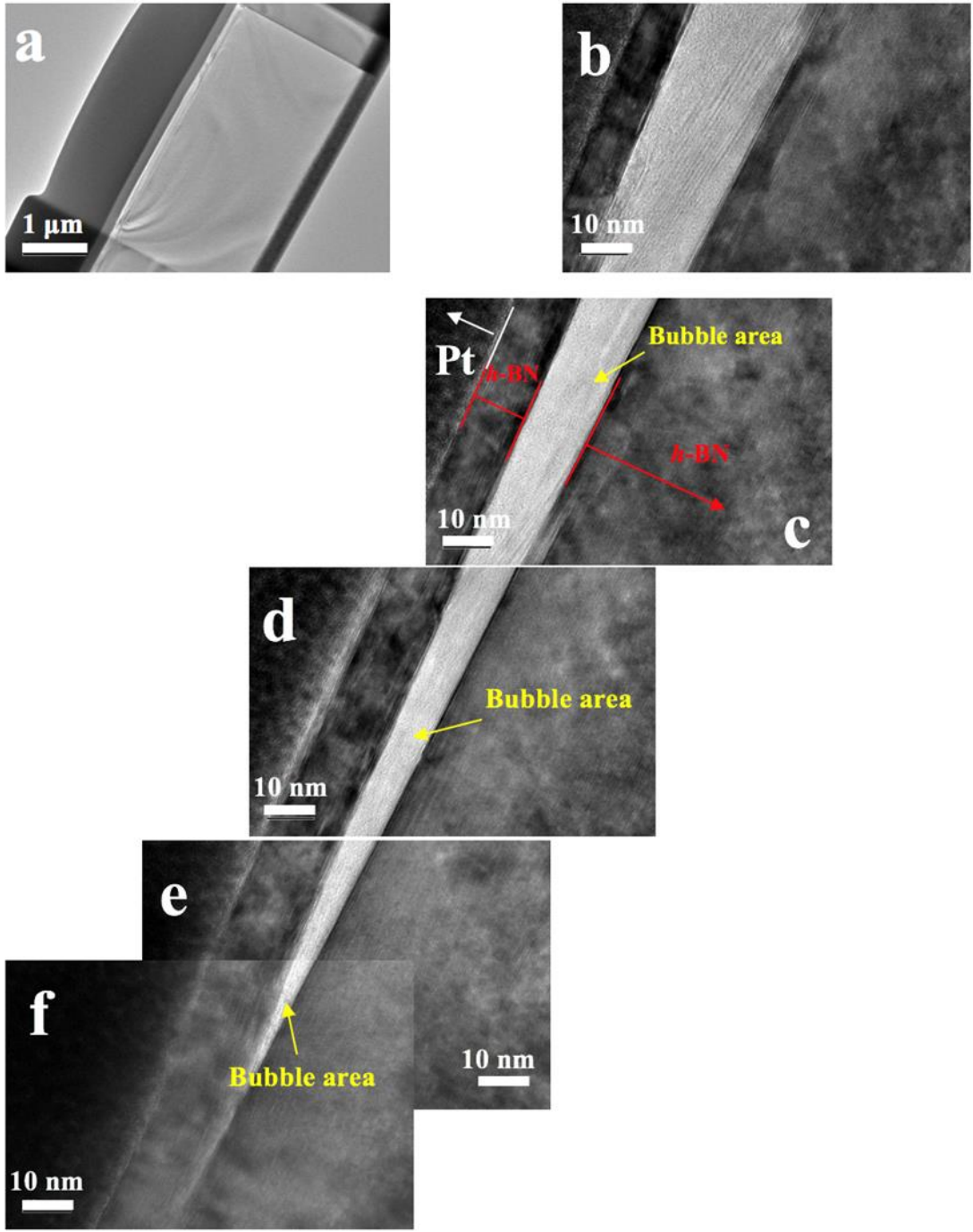

**Supplementary Figure 3 | Enlarged cross-sectional TEM view of bubble position inside *h*-BN. a,** Cross-sectional picture of the *h*-BN bubble-sample after FIB fabrication; **b-f,** Specific enlarged TEM images from middle to border area of *h*-BN bubble-position.

**Transmission Electron Microscopy (TEM) measurement:**

A group of cross sectional images measured by scanning electron microscope (SEM) and TEM are presented to characterize the *h*-BN bubble structure. The specimen for TEM measurement was prepared by Focused Ion Beam (FIB) system (Helios NanoLab

600) which is equipped with a function of SEM. Supplementary Fig. 2 shows the specific process to make a specimen of *h*-BN bubbles. The green rectangle indicated in Supplementary Fig. 2c is the area selected for e-beam induced Pt deposition. After Pt deposition (Supplementary Fig. 2e), the TEM specimen for bubble cross-sectional imaging was ready after FIB shaping (Supplementary Fig. 2f-g, Supplementary Fig. 3a). The SEM images are captured by the SEM in the FIB system. Then, the specimen was transferred to a TEM chamber (JEOL 2100F, operated at 200 kV). The enlarged cross-sectional scanning images by TEM are shown in a set of pictures which are manually combined to exhibit the *h*-BN bubble from the middle to the border area (Supplementary Fig. 2b-f). Note that the gap of the bubble is so narrow (in a few tens of nm) that we can hardly see it in the overall cross-sectional TEM image of Supplementary Fig. 2a.

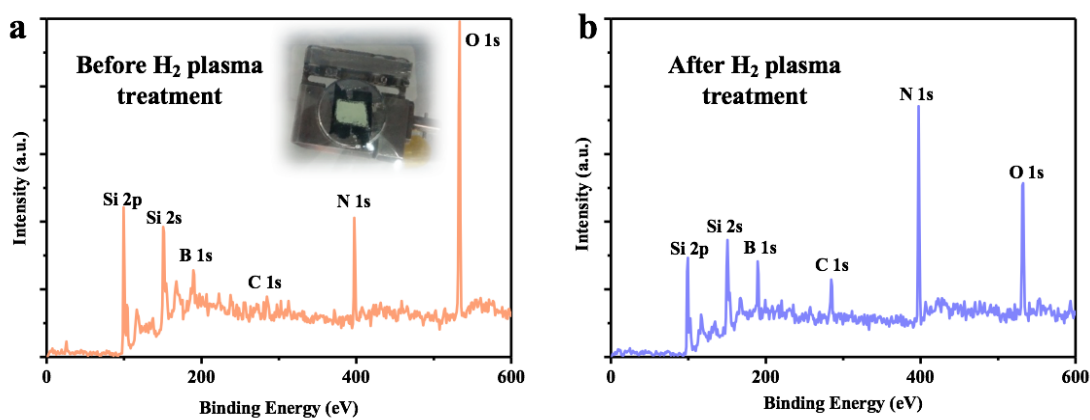

**Supplementary Figure 4 | XPS survey analysis of *h*-BN flakes on quartz substrate before (a) and after (b) $H_2$ plasma treatment in the full energy range.** The B 1s and N 1s peaks are visible. Inset shows a quartz substrate with *h*-BN flakes placed on a sample holder.

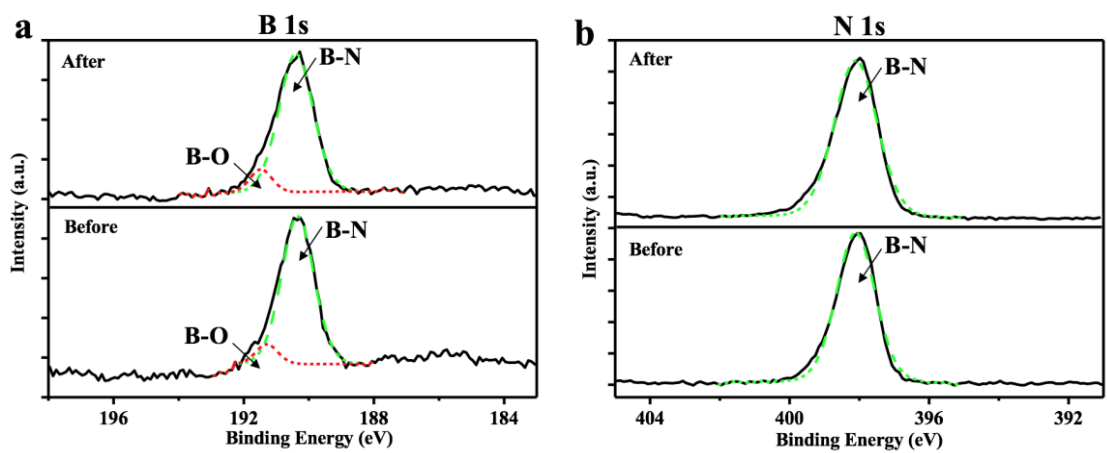

**Supplementary Figure 5 | Fine XPS spectra survey of *h*-BN flakes before and after H₂ plasma treatment**. B 1s (a) and N 1s (b) spectra obtained from *h*-BNs flakes. Dashed lines are fitting curves.

|  | B-N(FWHM) eV | B-O(FWHM) eV | N-1s eV |
|---|---|---|---|
| h-BN with bubbles (after plasma treatment) | 190.4(1.24) | 191.5(0.8) | 398.1(1.44) |
| h-BN (before plasma treatment) | 190.3(1.17) | 191.3(0.8) | 398.1(1.29) |

**Supplementary Table 1 | Peak position and FWHM of XPS spectra of the *h*-BN flakes before and after plasma treatment.**

**X-ray photoelectron spectroscopy (XPS):**

Firstly, X-ray photoelectron spectroscopy (XPS) measurement was carried out to investigate the influence of H₂ plasma treatment on *h*-BN flakes. Substrates with *h*-BN flakes are measured on SPECS XPS system using monochromatic Mg Kα line at a base pressure of $10^{-9}$ mbar before and after plasma treatment. The XPS survey analysis in the full energy range is shown in Supplementary Fig. 4. XPS spectra of *h*-BN flakes are calibrated with reference to C 1s at 284.5 eV. The specific measurement results are given in Supplementary Fig. 5. Supplementary Fig. 5a shows the narrow scan B 1s of *h*-BN before and after plasma treatment. As shown in Supplementary Fig. 5a, there is an obvious peak in the binding energy range from 184 eV to 196 eV, which can be fitted into two peaks. The main peak (the green dashed curve) corresponds to B-N bond of B 1s, while the tiny peak (the red dashed curve) corresponds to the B-O bond which were caused by annealing in O₂ flow at 600 ℃. Annealing in O₂ flow is necessary to remove tape residues after exfoliation. Supplementary Fig. 5b shows the N 1s core level spectra that were fitted to only one curve which corresponds to B-N bond. It is clear that the XPS spectra of *h*-BN flakes does not change obviously even after H₂ plasma treatment. As shown in Supplementary Table 1, the *h*-BN XPS peaks with bubbles have broadened, this phenomenon can be reasonably attributed to the *h*-BN expansion caused by bubbles formation.

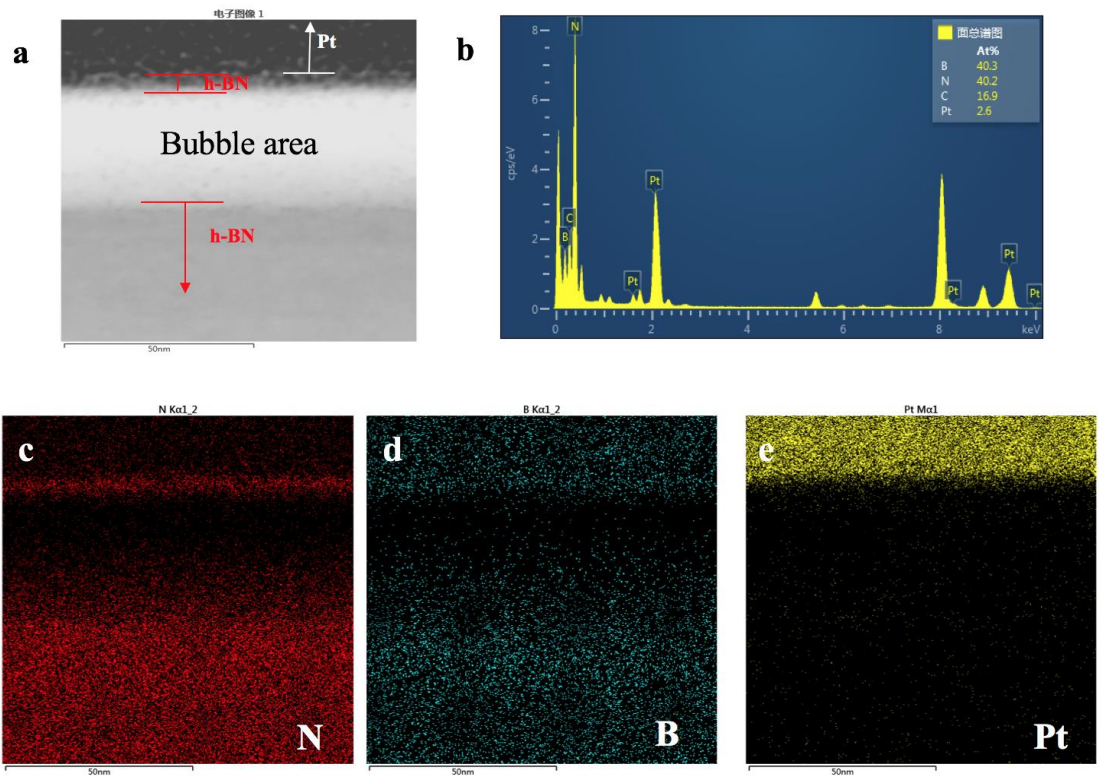

**Supplementary Figure 6 | EDX elemental analysis of the cross-sectional interface of *h*-BN specimen. a,** Scanning transmission electron microscope (STEM) image shows the selected EDX inspection area. *h*-BN and Pt area are indicated. **b,** The resulting spectrum and tabulated results reveal that B, N, and Pt are the main elements present with C element being mainly caused by hydrocarbon contamination. **c-e,** The corresponding elemental mapping of N, B and Pt, respectively. EDX elemental mapping indicates that the bubble area is empty.

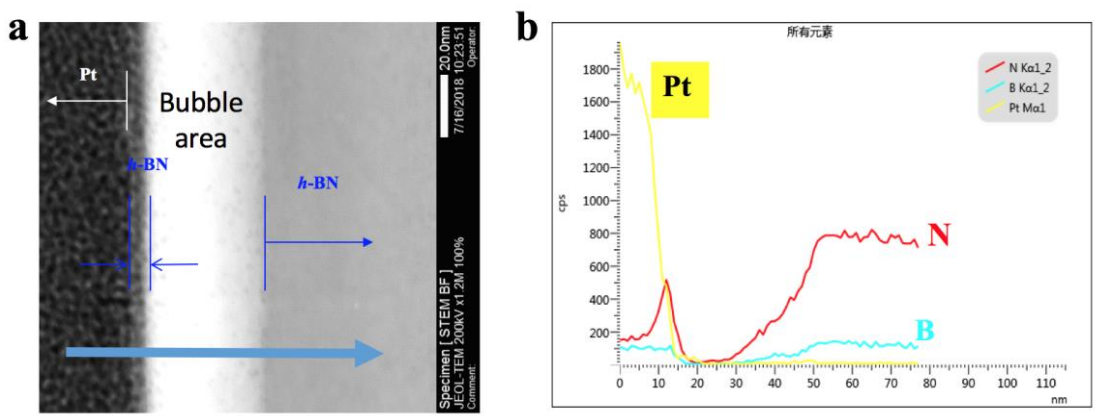

**Supplementary Figure 7 | Element line profile scan for Pt, N and B using EDX on STEM. a,** High-resolution STEM cross-sectional image of a *h*-BN bubble. **b,** The lateral distribution of Pt, N and B by using EDX analysis via line scan.

**Energy Dispersive X-Ray (EDX) analysis on elemental composition of *h*-BN:**

EDX is a semi-quantitative x-ray technique used to identify the elemental composition of materials. The scanning transmission electron microscope (STEM) is equipped with a EDX detector. Supplementary Fig. 6 shows EDX elemental analysis of the cross-sectional interface of *h*-BN specimen. As shown in Supplementary Fig. 6b, EDX spectrum shows peaks corresponding to the elements (B, N, C and Pt), where B and N the main elements. Elemental mapping of a sample and image analysis are given in Supplementary Fig. 6c-e. EDX elemental mapping clearly shows that the bubble area is empty.

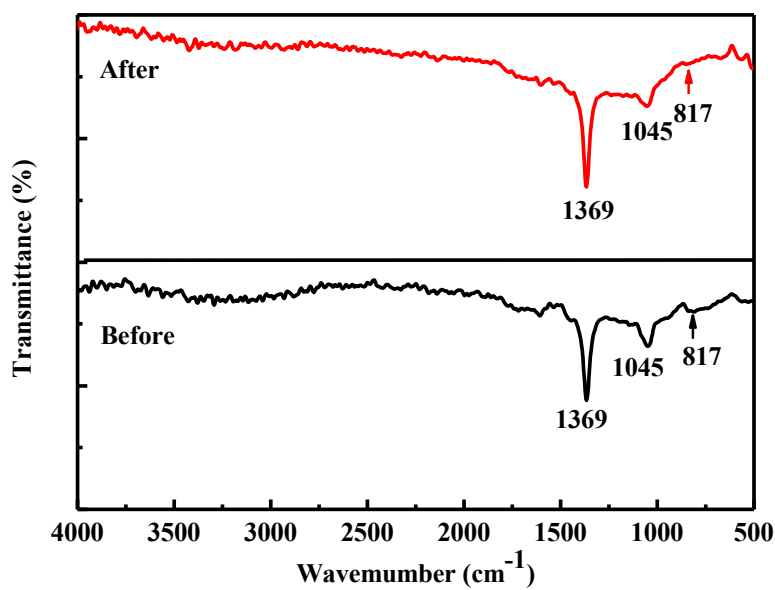

**Supplementary Figure 8 | FTIR spectra of the *h*-BN flakes on silicon substrates before and after H$_2$ plasma treatment.**

**Fourier-transform infrared (FTIR) spectra:**

Hydrogen plasma treatment was carried out on *h*-BN flakes exfoliated on pristine <100> silicon substrate to produce bubbles. FTIR measurement was conducted at ~1 mbar (on Bruker IFS 66v/S) before and after hydrogen plasma treatment. The FTIR spectra of the *h*-BN before and after H$_2$ plasma treatment are shown in Supplementary Fig. 8. There are 3 obvious dips in the range from 500 to 4000 cm$^{-1}$. The dip at 1369 cm$^{-1}$ was derived from the in-plane stretching vibration of *h*-BN, and the dip at 817cm$^{-1}$ was owning to out-of-plane bending absorption of the B-N-B. The absorption peak appears at 1045 cm$^{-1}$ was a fingerprint of the presence of Si-O bonds, which was derived from the substrate. The similar characteristics in FTIR spectra indicate that the plasma treatment did not chemically cause obvious change in *h*-BN flakes.

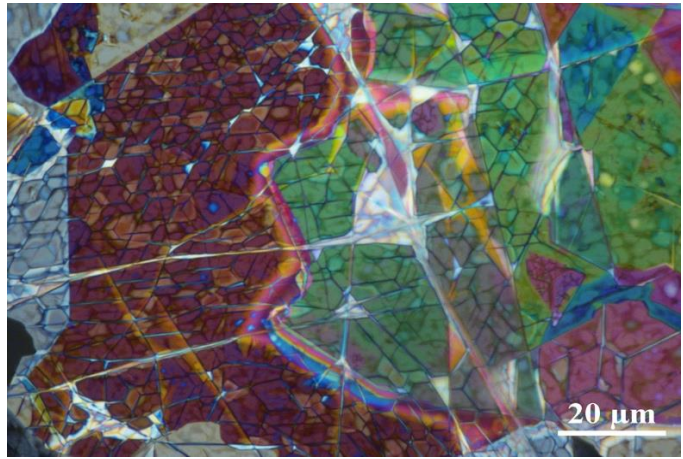

**Supplementary Figure 9 | Optical microscope image of a *h*-BN flake with dense bubbles on a quartz substrate, which is the sample prepared for the mass spectra measurement.**

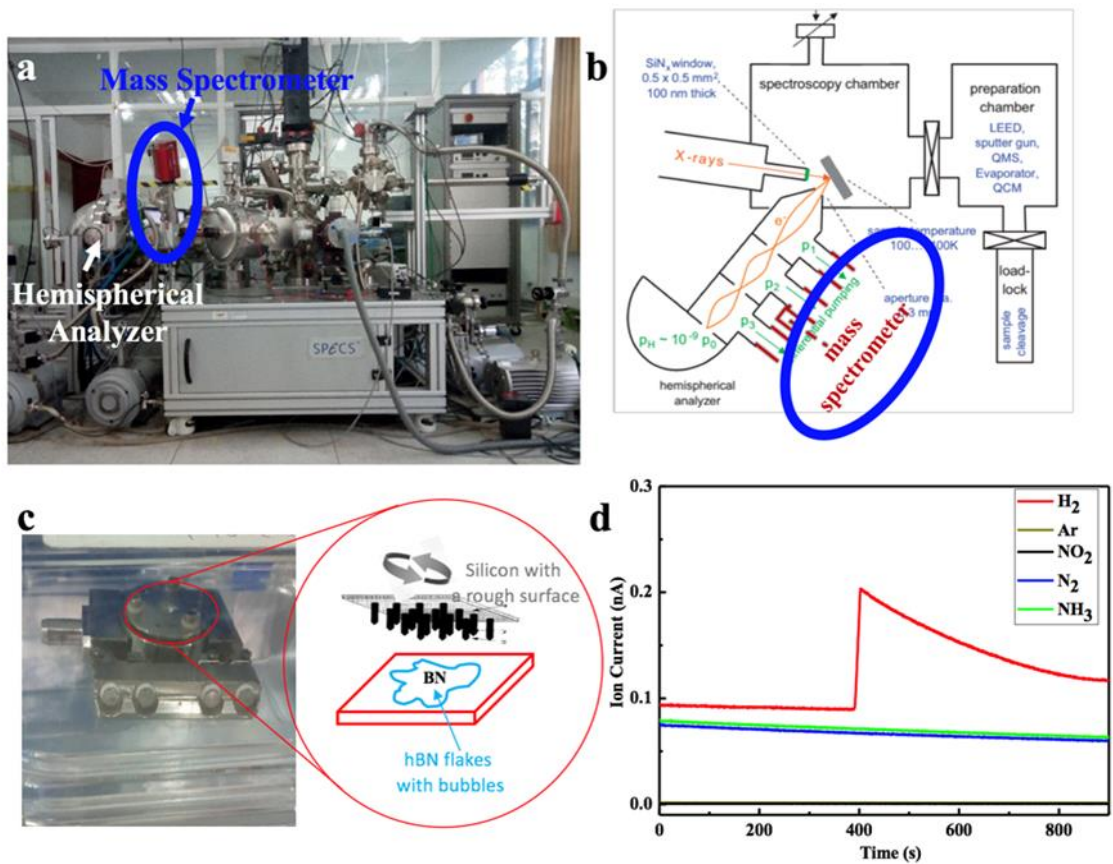

**Supplementary Figure 10 | Experimental details about the mass spectra analysis on *h*-BN bubbles. a,** A UHV system equipped with a mass spectrometer. **b,** Schematic for the vacuum system. **c,** sample holder used for the mass spectrum analysis. **d,** Real-time monitoring on different gases by mass spectrometer.

**Mass spectra (MS) analysis:**

Mass spectrometry measurement can be conducted to analyze gases in the bubble. Mass spectra (MS) analysis was carried out on a UHV vacuum system with quadrupole mass spectrometer (Pfeiffer QMS 220 M). The sample for MS measurement is a quartz substrate in size of 10 × 10 mm with a high density of $h$-BN flakes, which has very dense bubbles (Supplementary Fig. 9). The measurement system is shown in Supplementary Fig. 10a and the schematic of its structure is given in Supplementary Fig. 10b. The substrate with $h$-BN bubbles were installed in a homemade sample holder (see Supplementary Fig. 10c). As shown in Supplementary Fig. 10c, another silicon substrate with rough surface were placed on the quartz substrate with $h$-BN bubbles surface, and both substrates were assembled on sample holder gently. The holder has a screw which can slide the top silicon substrate. The sliding can spoil the $h$-BN bubbles on underlying quartz substrate. The experiment was carried out in a UHV chamber with a base pressure of $10^{-9}$ mbar. After the base pressure in the chamber became stable, we rotated the screw to break $h$-BN bubbles. The mass spectrum variation was recorded and plotted in Supplementary Fig. 10d. We set several possible gases ($H_2$, Ar, $NO_2$, $N_2$ and $NH_3$) for monitoring. As shown in Supplementary Fig. 10d, only the hydrogen increased suddenly after the bubbles were broken, and then gradually decreases with time. This experiment proves that the gas inside the bubble is hydrogen.

Here, we discuss some details about the experimental design. The density of $h$-BN flakes on 10×10 mm quartz substrate is about 30×30 pieces, and the density of bubbles on each piece of $h$-BN flake is about 100×100. Assuming the pressure in a bubble is about 10 bar, and the volume of the bubble is about 10 μm × 10 μm × 0.1 μm. The volume of the chamber is 0.1 m$^3$. Even if 1% of bubbles were broken, the pressure will reach about $9 \times 10^{-8}$ mbar. Compared with the base pressure of chamber, $10^{-9}$ mbar, the gas is sufficient to be detected.

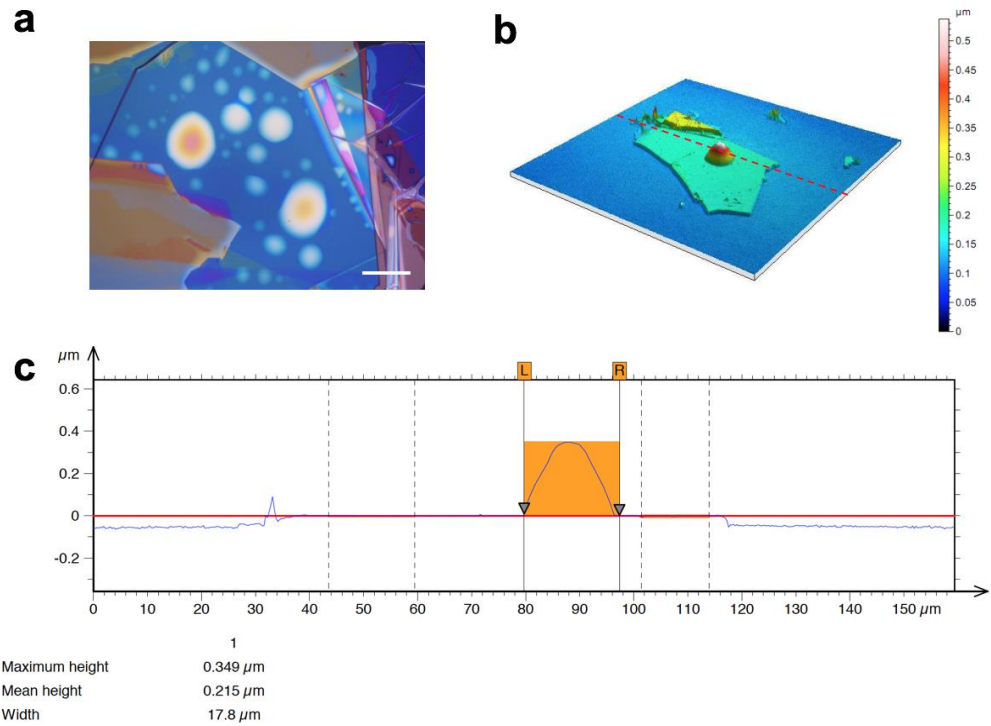

**Supplementary Figure 11 | Large *h*-BN bubbles obtained after a 500-minute H-plasma treatment. a**, Optical image of the bubbles produced on *h*-BN after a 500-minutetreatment with H-plasma. Scale bar: 20 μm. **b**, Topographic image of a hydrogen bubble (~17.8 μm in diameter) on *h*-BN captured by 3D laser confocal microscopy (NanoFocus usurf). **c**, Height profile along the red dashed line shown in (**b**), revealing a height of 349 nm at the peak of the bubble.

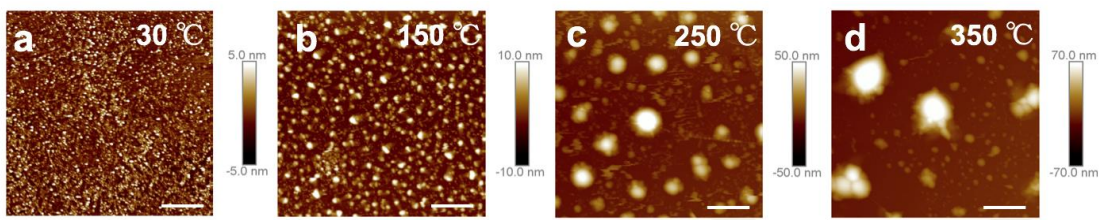

**Supplementary Figure 12 | AFM images of hydrogen bubbles after H-plasma treatment.** The sample temperature was set at **a**, 30 ℃, **b**, 150 ℃, **c**, 250 ℃ and **d**, 350 °C. Scale bars: 2 μm.

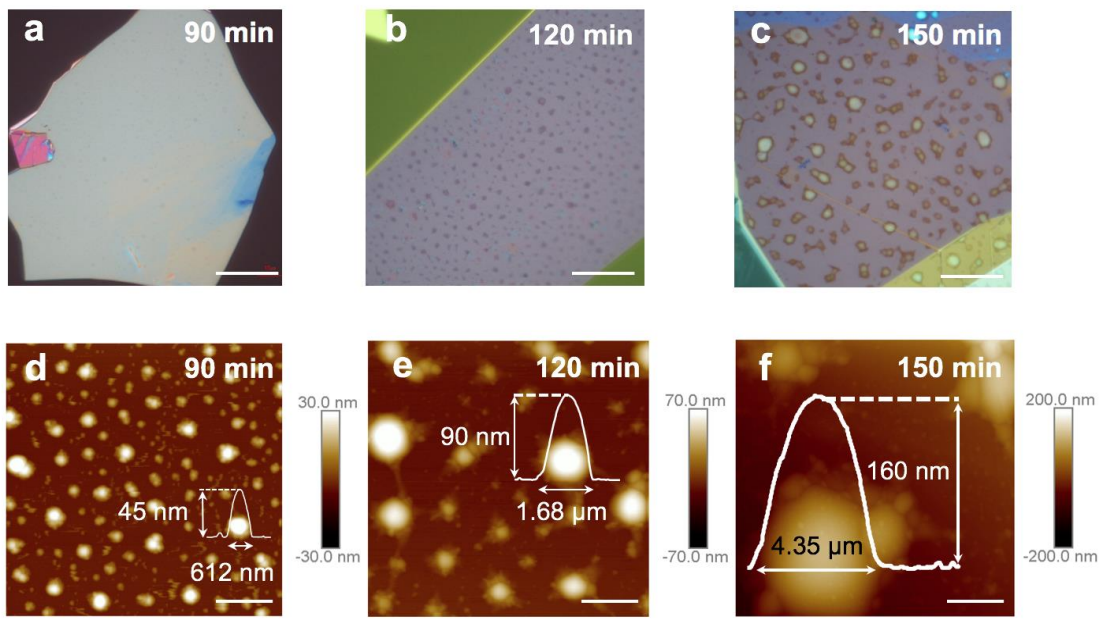

**Supplementary Figure 13 | Influence of the H-plasma treatment duration on the bubble dimensions. a-c**, Optical images of *h*-BN flakes obtained after H-plasma treatment for 90, 120 and 150 minutes, respectively. Scale bars: 20 μm; **d-f**, AFM height images of the samples shown in **(a-c)**, respectively. The profiles in **(d-f)** give detailed information about the height and diameter of the selected bubbles. Scale bar: 2 μm.

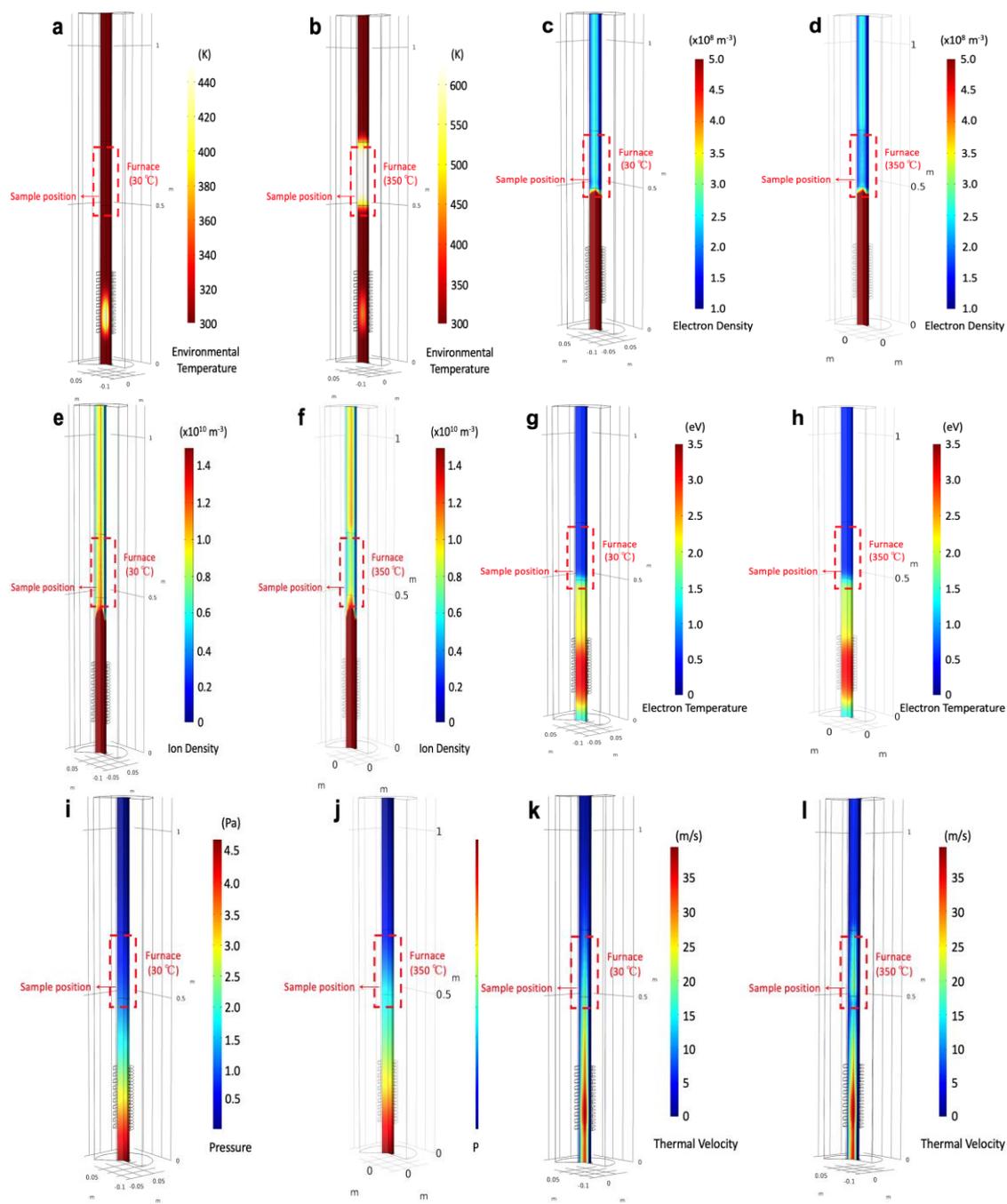

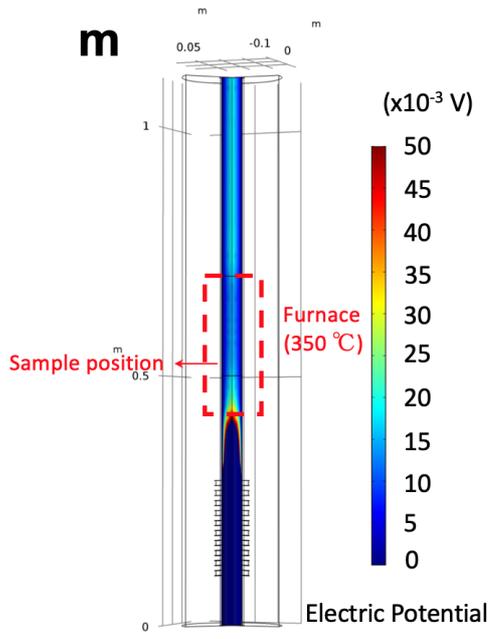
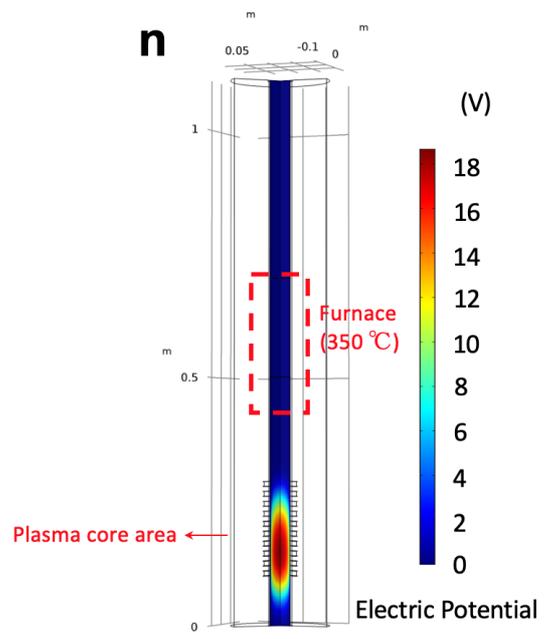
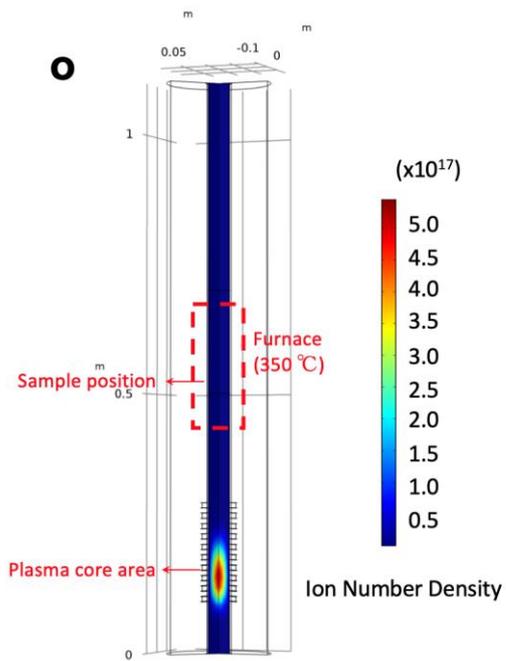
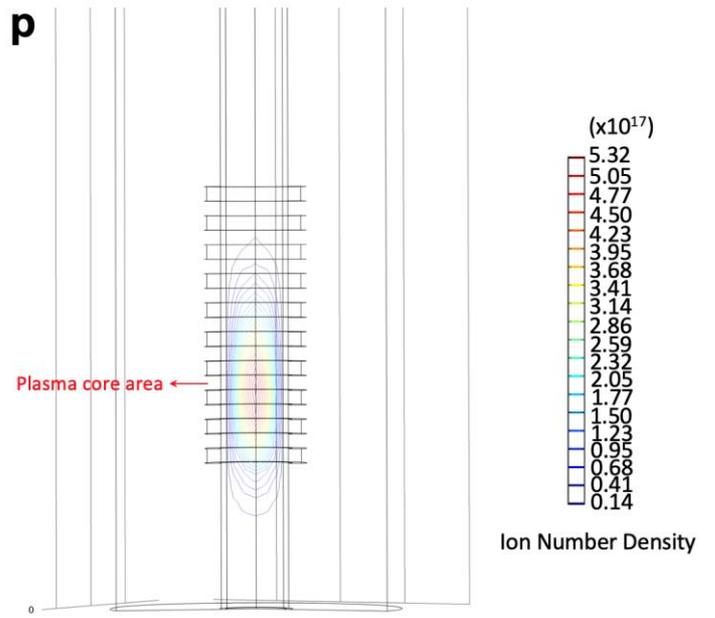

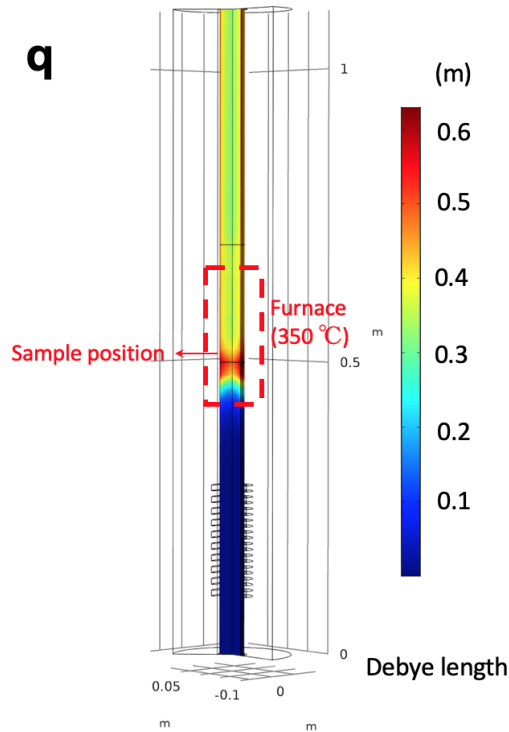

**Supplementary Figure 14 |Distribution of different physical parameters in the tube under different conditions.** The environmental temperature distribution in the tube when the temperature is set from **a,** 30 ℃ to **b,** 350 ℃ in the furnace marked as a dashed frame. Contrast of electron density in the tube when the furnace temperature is set from **c,** 30 ℃ to **d,** 350 ℃. Contrast of ion density in the tube when the furnace temperature is changed from **e,** 30 ℃ to **f,** 350 ℃. Contrast of electron temperature in the tube when the furnace temperature is set from **g,** 30 ℃ to **h,** 350 ℃. Contrast of pressure in the tube when the furnace temperature is set from **i,** 30 ℃ to **j,** 350 ℃; and Contrast of thermal velocity in the tube when the furnace temperature is changed from **k,** 30 ℃ to **l**, 350 ℃. **m,** Details of the electric potential distribution at the sample position when the furnace temperature is set to 350 ℃. **n,** Details of the electric potential distribution at plasma core area (there is no heater at plasma core region). **o,** Details of the ion number density distribution at the plasma core region. **p,** Zoom-in contour of electric potential at the plasma core region. **q,** Distribution of Debye length in the whole tube when the furnace temperature is set at 350 ℃.

We try to do detail analysis about plasma driven kinetic mechanisms. Supplementary Fig. 14 shows the collateral effects of some different physical parameters result from the variation of the sample temperature by Comsol simulations. The simulations exhibit some differences of physical distribution in the plasma tube when the environmental temperature at sample position was set to 30 ℃ and 350 ℃ by the furnace (Supplementary Fig. 14a-b). It is obvious that when the environmental temperature increases, the electron density is decreased along with the ion density (Supplementary Fig. 14c-d, e-f), the electron temperature is slightly decreased but almost keep the same (Supplementary Fig. 14g-h), and the pressure and thermal velocity at the sample position are both increased (Supplementary Fig. 14i-j, k-l). From the simulation results, we found that the ion density decreases while the thermal dynamic velocity of ions increases. This is a possible reason to explain why the density of bubble decreases and the size of bubbles increases when the samples were heated up (See in Fig. 3c in main manuscript). Both Supplementary Fig. 14m and Supplementary Fig. 14n show the distribution of electric potential in the reaction tube from the same simulation. As there is a large change in the magnitude of electric potential between the plasma core region and the sample area, we plot the electric distribution of these two areas separately in different scale for clarity. Supplementary Fig. 14o presents the ion number density at the plasma core area. To clearly show the ion number density at the plasma core area, we also put a zoom-in contour view in Supplementary Fig. 14p. The distribution of Debye length is also simulated, and presented in Supplementary Fig. 14q. Debye length reaches its maximum at the sample position (~ 0.5 m). This value is much higher than the diameter of our reaction tube (0.042 m). It indicates that the quasi-neutrality at the sample position is destroyed. This interprets why the electron density is 2 orders lower than the ion density at the sample position in our pervious simulation. Moreover, a higher value of Debye length in the sample position can be achieved if the heated furnace is set at 350 ℃. Normally, Debye length depends on density of ion number and environmental temperature. Debye length increases when ion number density decreases or when the environmental temperature increases.

**Study of bubble-formation-distribution at different sample areas in reaction tube.**

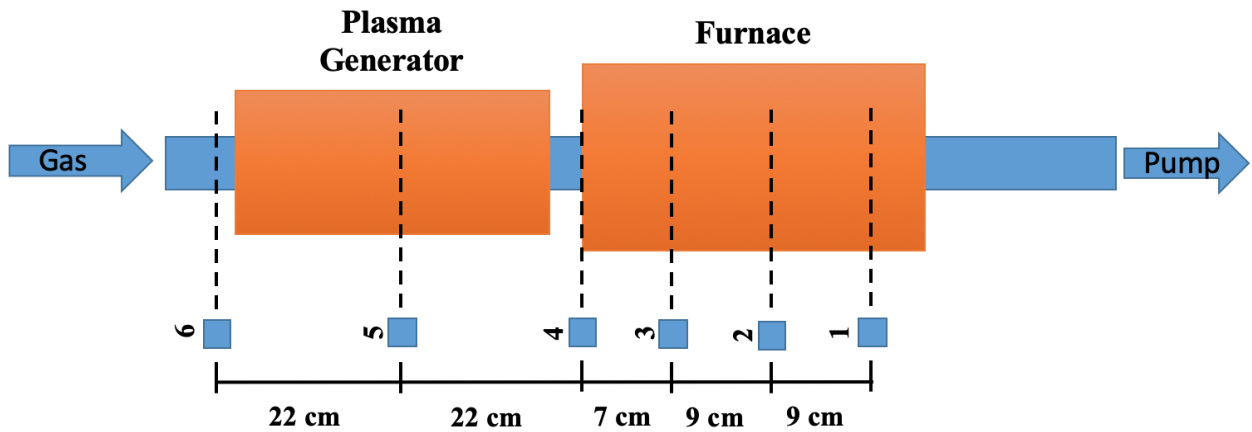

**Supplementary Figure 15 | Schematic of different positions to place the *h*-BN samples along the tube. (from Position 1 to 6)**

**Position-1**

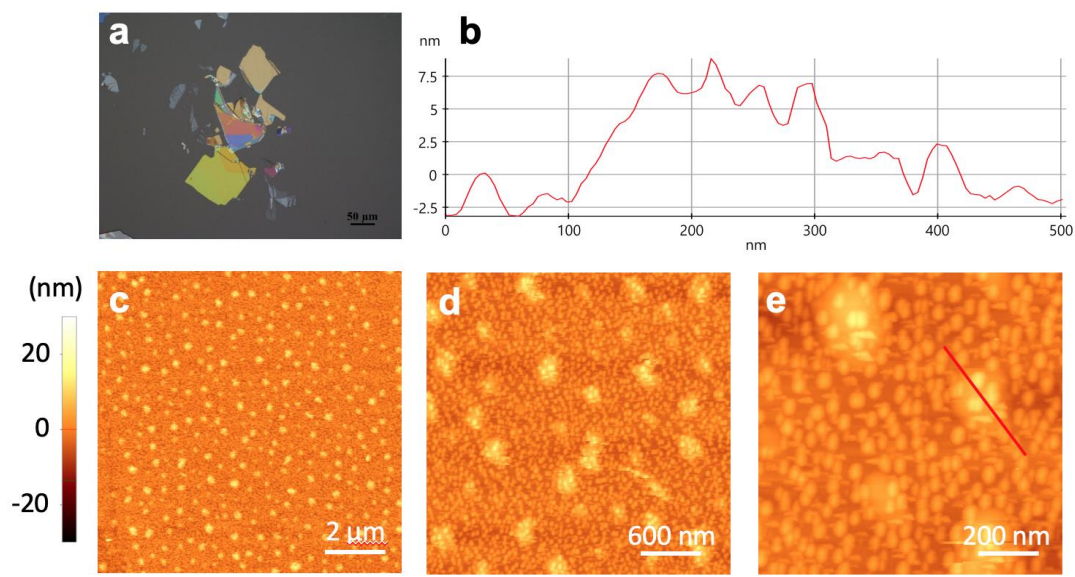

## Position-2

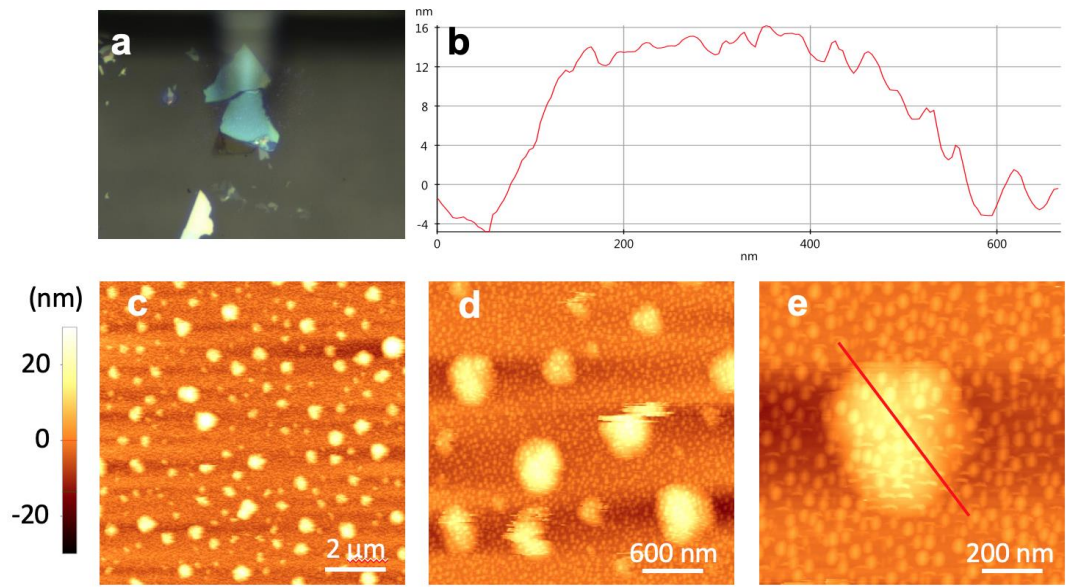

## Position-3

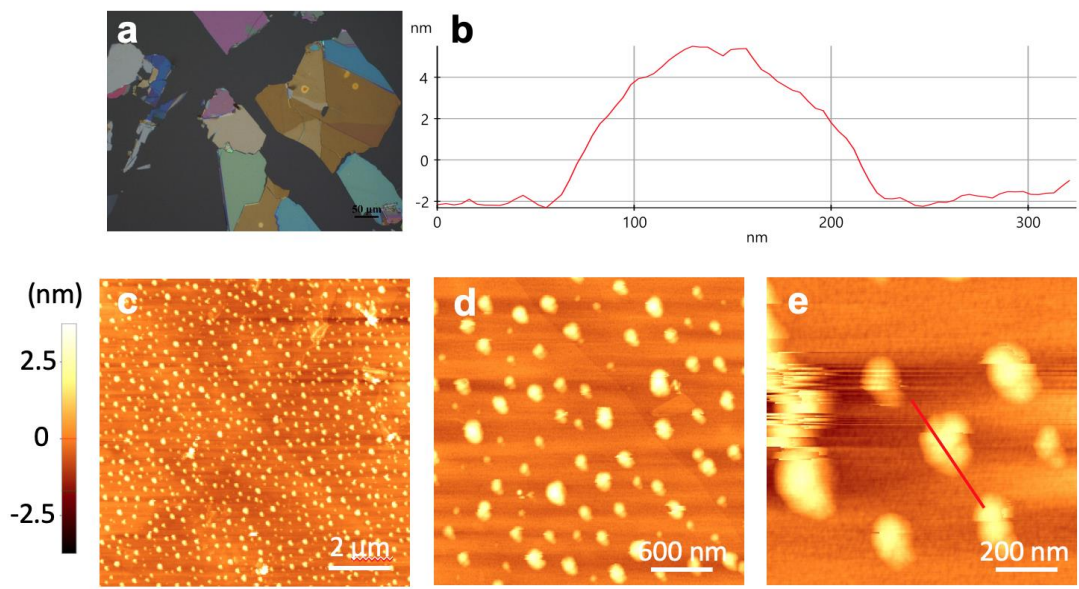

## Position-4

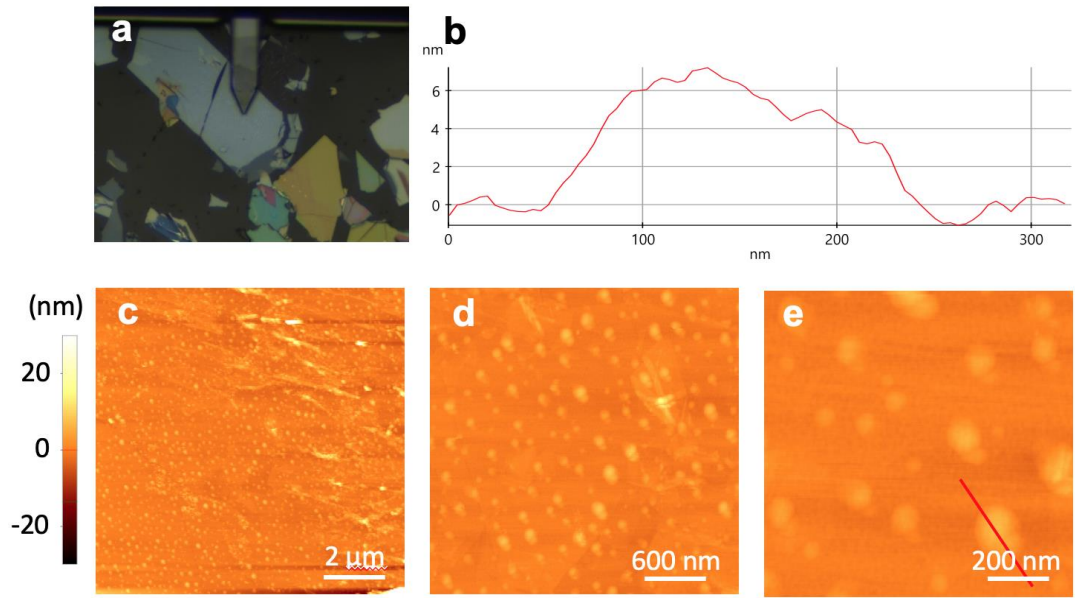

## Position-5

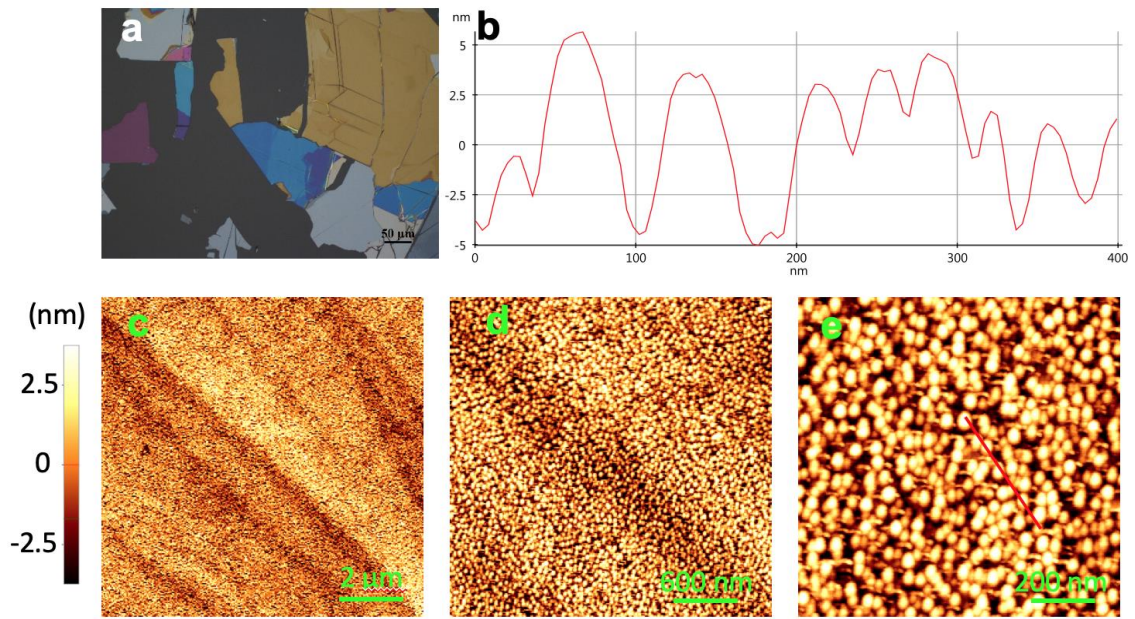

**Position-6**

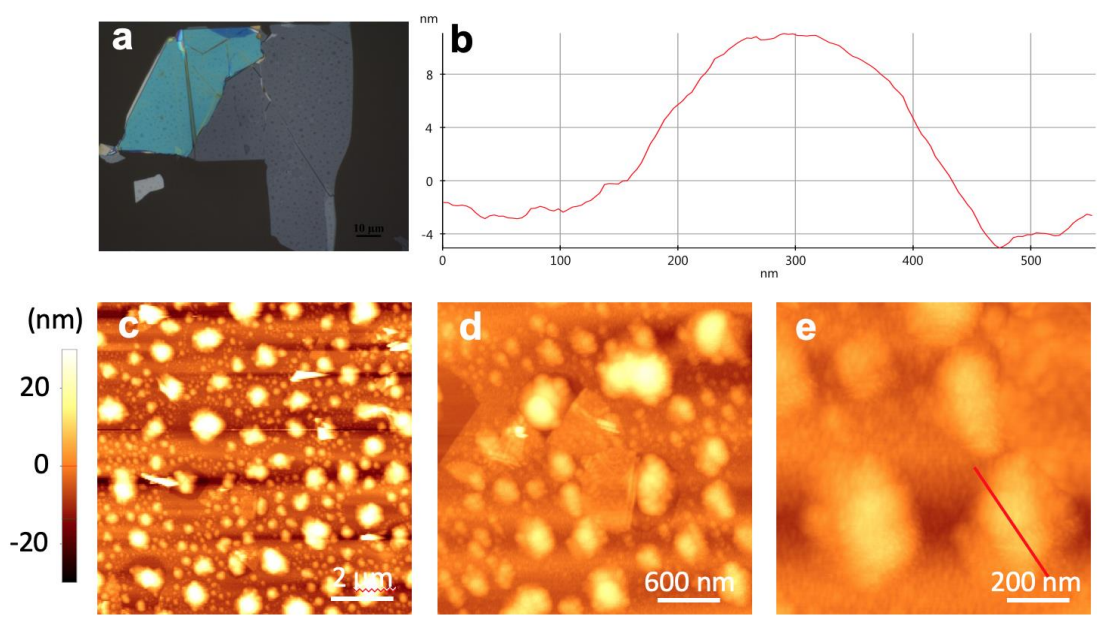

**Supplementary Figure 16 | Optical image and AFM topographic image of *h*-BN samples placed from Position 1 to Position 6.** For the *h*-BN samples at each position, **a,** The optical image of *h*-BN sample. **b,** The cross-sectional profile which is corresponding to the red line marked in (e). **c-e,** AFM topographic images in different scan area.

We carried out the experiments accordingly. We placed six *h*-BN samples at Position 1-6 along in the tube as shown in Supplementary Fig. 15 in order to investigate the difference in bubble formation. There are detailed experimental parameters: RF power in 100 W, the center of the furnace at 350 ℃, a $H_2$ flow in a rate of 3 sccm, a pressure in 3 Pa and a duration of 60 minute for plasma treatment. AFM and optical images of the *h*-BN samples were taken and the detailed results are shown in Supplementary Fig. 16. As shown in Supplementary Fig. 16, the bubbles on *h*-BN sample at Position 2 (furnace center, 350 ℃) and Position-6 (at the front of the RF coil) are much larger than those on other samples while there is almost no big bubble visible on *h*-BN placed at Position 5 (plasma core area near the center of the RF coil) under optical microscope. However, it is found that there are masses of tiny bubbles in a very high density on *h*-BN samples placed at Position 5 by AFM scanning. To understand these phenomena, we carried out the distribution simulation of both the electric potential (Supplementary Fig. 14m-n) and the ion number density (Fig. Supplementary Fig. 14o-p) in the tube. At the center of the RF coil (position 5), the electric potential reaches its maximum but its gradient is much lower than that of the entrance area of RF coil (Position 6). This difference could lead to the distinction of the ion injection quantity on these two samples (the height of bubbles on *h*-BN at Position 5 is averagely < 5 nm. But for sample placed at Position 6, that is > 15 nm). Moreover, the plasma core region at Position 5 also possesses the maximal ion number density, according to Supplementary Fig. 14o-p. And therefore, it is reasonable that the *h*-BN flakes at Position 5 has the highest density of the bubbles among all *h*-BN samples. On the other hand, *h*-BN samples located at Position 1-4 also show different distributions of the bubble formation. At the center of the furnace, both the height and diameter values of bubbles reach their maximum of ~18 nm and ~600 nm at Position-2 while bubbles on *h*-BN at Position 1,3 and 4 could only be raised to ~7-10 nm high with the average diameter of ~100-300 nm. It is likely that the environmental temperature of 350 ℃ at the center of the furnace offers some additional energy to the ions, which makes them easier to penetrate the *h*-BN mesh and gives rise to the redistribution and size changing of larger bubble formation at Position 2.

Generally, the energy of the accelerated ions is determined by the comprehensive influence from ion density, Debye length and potential gradient in plasma. According to the analysis in Reply to Comment 3, the higher density of ions may remarkably increase their collision probability with other ions and electrons, and then increase the probability of their recombination to hydrogen molecules. It could markedly shorten the mean free path of ions at the plasma core area and restrict the ions to obtain enough energy for their accelerations. As such, the density of bubbles in the plasma core area should be very high and exhibits relatively small size. Comparatively, ions can get more energy at the sample area due to the lower ion number density and finally give rise to thin but relatively large bubbles. As shown in Supplementary Fig. 16, the experimental results are consistent with our analysis above: Bubbles in the **plasma core area** are small but dense. while bubbles **at other area** show relatively large but thin.

**Low temperature AFM measurement of the bubbles**

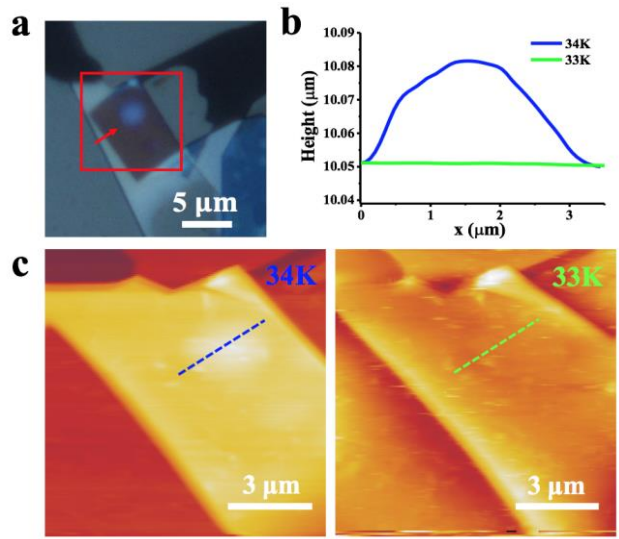

**Supplementary Figure 17 (A) |** The vanished bubble on *h*-BN surfaces when the temperature is cooling down from 34 K to 33 K.

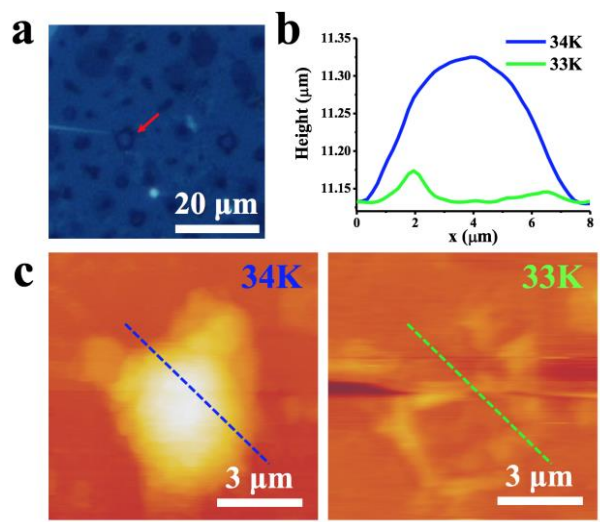

**Supplementary Figure 17 (B) |** The vanished bubble on *h*-BN surfaces when the temperature is cooling down from 34 K to 33 K.

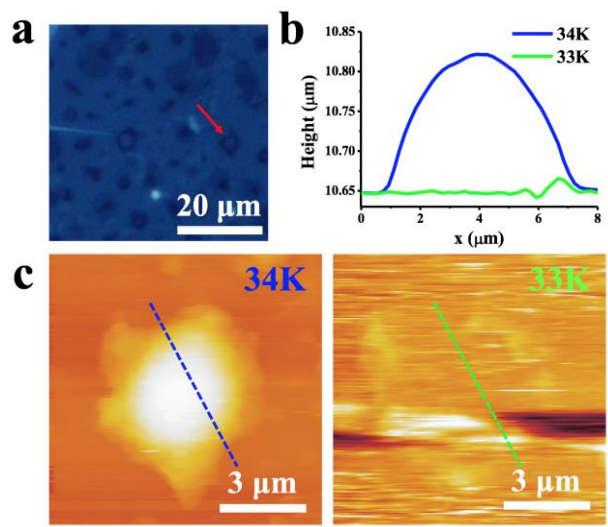

**Supplementary Figure 17 (C) | The vanished bubble on *h*-BN surfaces when the temperature is cooling down from 34 K to 33 K.**

During the cooling process from room temperature, we found that bubbles shrink gradually. It can be easily understood by the second law of thermodynamics: $PV/T = constant$, where pressure inside the bubble ($P$) mainly depends on the elastic modulus of *h*-BN. It is known that elastic modulus always keeps constant even when the temperature varies. As a result, the volume of the bubble will decrease during the cooling process. The formation of wrinkles on *h*-BN surfaces shown in Supplementary Fig. 17(A-C)c may be related to the shrinking of the bubbles.

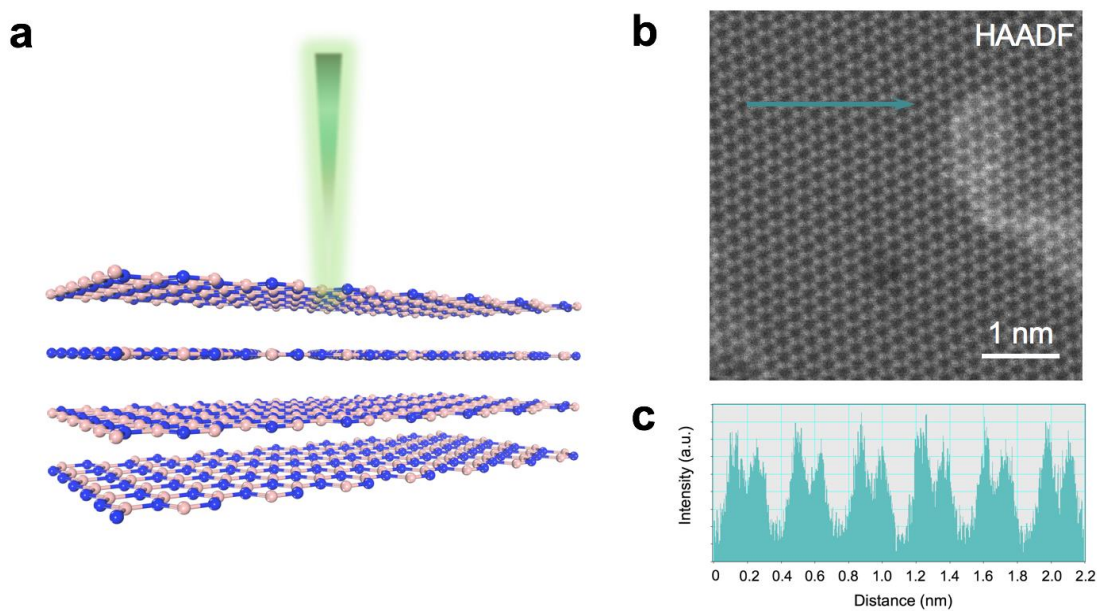

**Supplementary Figure 18 | Plane-view STEM measurement for a *h*-BN multilayer.
a,** Schematic of how the STEM electron beam is applied onto the *h*-BN sample. The electron beam penetrates the *h*-BN along its [0001] crystallographic plane. **b,** Plane-view HAADF image of the multilayer *h*-BN sample. **c,** Profile of intensity corresponding to the green arrow-area noted in **(a)**, shows AA′ stacking structure of the multilayer h-BN crystal. (if it were AA stacked, i.e. N above N and B above B, the two atomic columns would have a very different HAADF intensity; and any staggered layer arrangement would show up as intensity within the hexagons).

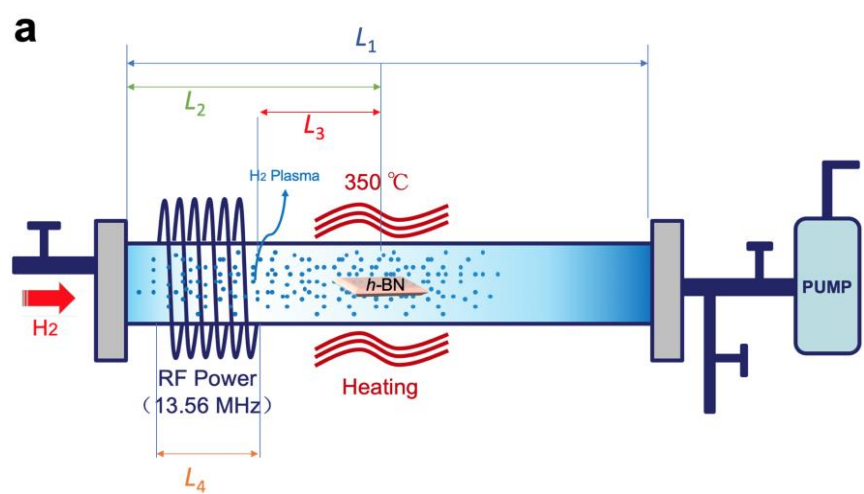

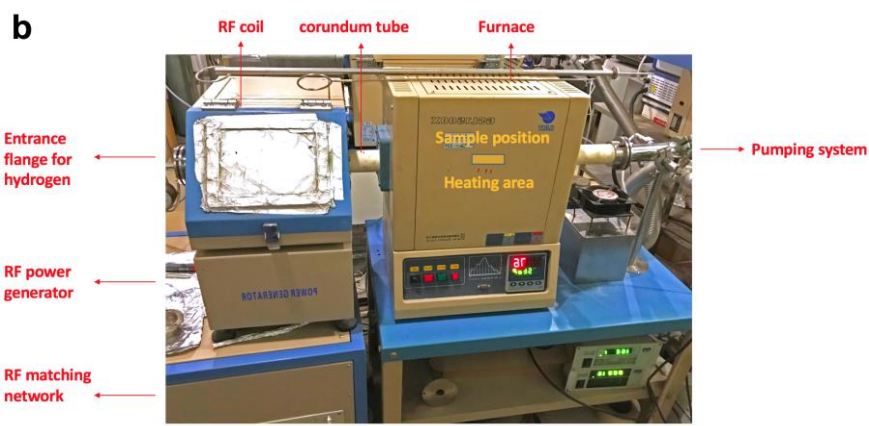

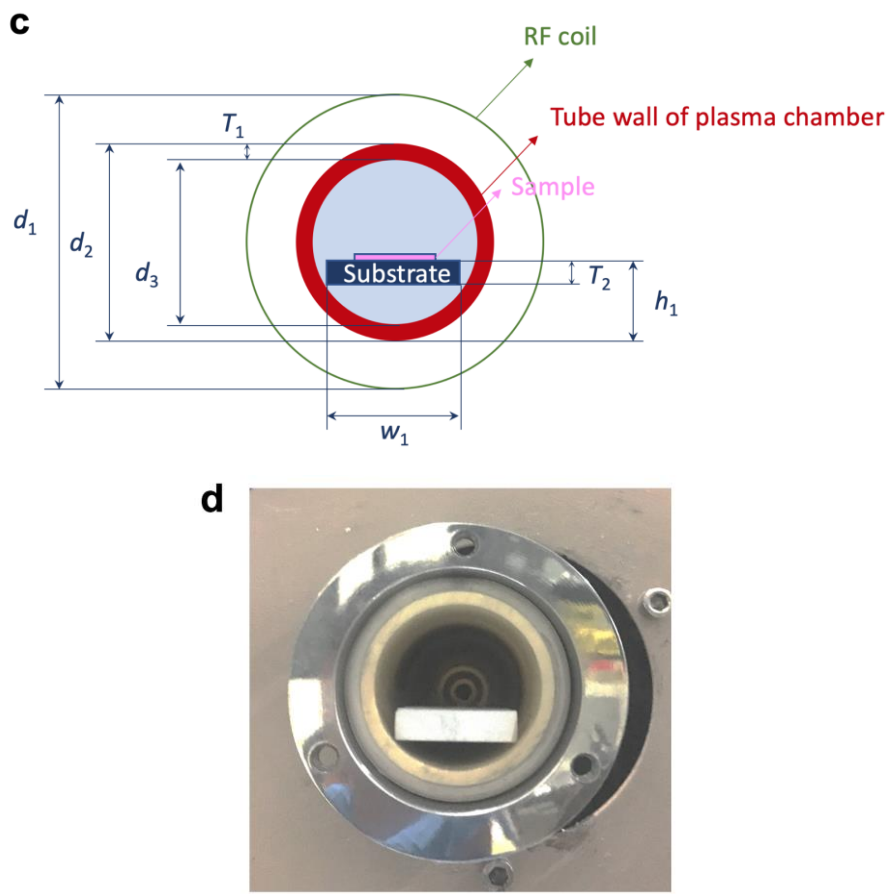

**Supplementary Figure 19 | Setup used to fabricate hydrogen bubbles on *h*-BN. a,** Side view of the furnace for bubble fabrication, where $L_1$ (110 cm) is the length of the plasma tube, $L_2$ (60 cm) is the distance from the entrance to the *h*-BN sample, $L_3$ (30 cm) is the horizontal distance between the RF coil and the *h*-BN sample and $L_4$ (20 cm) represents the length of the RF coil. **b,** Lateral photograph of the Plasma system corresponding to (**a**), captions in the image shows different functional units of the plasma system. **c,** Cross-sectional view of the plasma chamber system for bubble fabrication, where $d_1$ (12 cm) denotes the diameter of the RF coil, $d_2$ (5 cm) is the outer-wall diameter of the plasma tube, $d_3$ (4.2 cm) is the inner-wall diameter of the plasma tube, $T_1$ (0.8 cm) is the thickness of the wall of the plasma tube, $T_2$ (0.7 cm) is the thickness of the corundum substrate for sample (the *h*-BN sample is exfoliated onto a quartz substrate which is placed on the corundum substrate), $w_1$ (3.1 cm) is the width of the corundum substrate and finally $h_1$ (1.9 cm) represents the height of the sample from the bottom of the tube. **d,** Cross-sectional photograph of the entrance of plasma tube and the corundum sample holder.

**Boiling point and critical point of different gases**

|  | Boiling point | Critical point |  |
|---|---|---|---|
| **Hydrogen** | 20.271 K | 32.938 K, | 1.2858 MPa |
| **Helium** | 4.222 K | 5.1953 K, | 0.22746 MPa |
| **Nitrogen** | 77.355 K | 126.192 K, | 3.3958 MPa |
| **Oxygen** | 90.188 K | 154.581 K, | 5.043 MPa |
| **Argon** | 87.302 K | 150.687 K, | 4.863 MPa |
| **Methane** | 111.65 K | 190.6 K, | 4.64 MPa |

**Supplementary Table 2 | The boiling points and liquid-vapor critical points of different gasses adapted from Wikipedia.** [1]

References:
[1] Referred from *wikipedia.org*

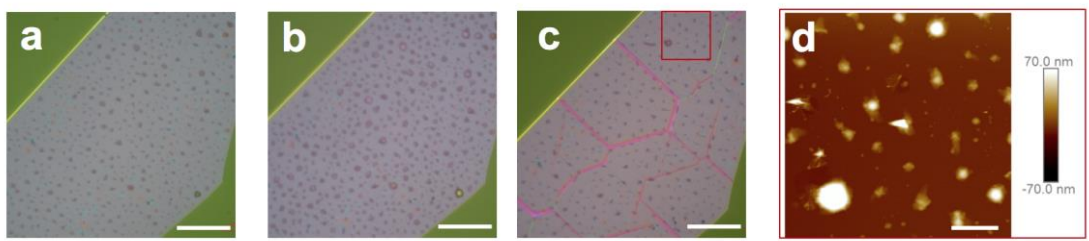

**Supplementary Figure 20 | Thermal stability study of *h*-BN bubbles with trapped hydrogen. a**, Optical image of *h*-BN bubbles produced by a 120-minute H-plasma (100 W) treatment at 350 ℃; the image was collected at a sample temperature of ~30 ℃. **b**, Optical image of the same *h*-BN flake taken at 300 ℃ on a heating stage under ambient air. The optical image shows swelling of the bubbles. **c**, Optical image of the same *h*-BN flake after annealing in an Ar/$O_2$ atmosphere at an 800 ℃; most bubbles are still visible after the high-temperature treatment, while *h*-BN wrinkles appear on the *h*-BN surface. **d**, AFM height image of the area marked by the red box in **(c)** showing that most bubbles survived after high-temperature annealing in an Ar/$O_2$ atmosphere. Scale bars: **(a-c)** 20 μm and **(d)** 4 μm.

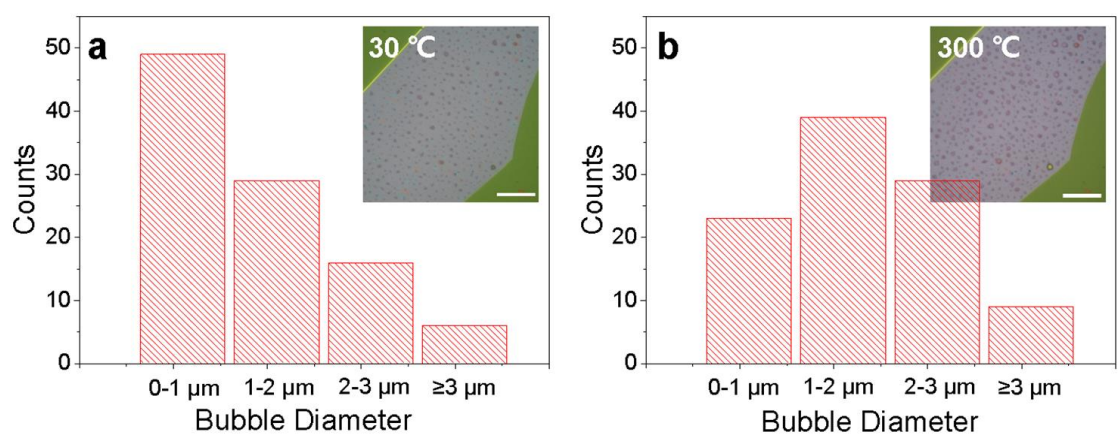

**Supplementary Figure 21 | Distribution of the bubble dimensions at different heating temperatures. a**, Statistics of the bubble diameter at 30 ℃; **b**, Size distribution of the *h*-BN microbubbles at 300 ℃. The insets in **(a, b)** show the optical images of the same *h*-BN flakes as those provided in Supplementary Fig. 20. The scale bars in the insets represent 20 μm.

**Bubble expansion effect on a heated substrate**

An *h*-BN sample with bubbles was placed on a heating stage, where the temperature could be controlled from 30 to 300 ℃. An optical microscope (Eclipse LV150, Nikon) was used to record variations in the bubble morphology for the same *h*-BN flake at both 30 ℃ and 300 ℃. Supplementary Fig. 21a exhibits the diameter distribution of the bubbles at 30 ℃, while Supplementary Fig. 21b shows the distribution of the bubble diameters on the same *h*-BN flake at 300 ℃. As shown in Supplementary Fig. 21, the bubble diameters were clearly enhanced when the substrate temperature was increased from 30 ℃ to 300 ℃. The enhancement of the bubble size reveals that the evolution in the bubble size during the heating process was dominated by the expansion of hydrogen gas, which can be attributed to the fact that the expansion of hydrogen molecules overcomes the fierce competition with the shrinkage of the *h*-BN lattice during heating of the substrate. We also noticed that the bubbles recovered the original dimensions after the substrate cooled to room temperature.

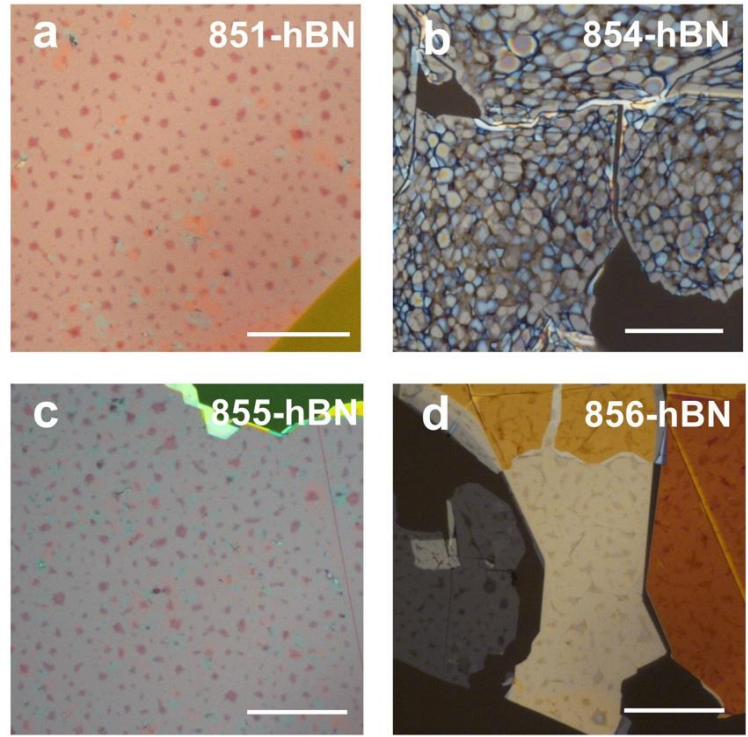

**Supplementary Figure 22 | Evaluation of the structural quality of *h*-BN flakes via H-plasma treatment.** Optical images of different *h*-BN samples labeled as: **a**, 851-*h*-BN, **b**, 852-*h*-BN, **c**, 853-*h*-BN and **d**, 855-*h*-BN after treatment with H-plasma for 120 minutes at 350 ℃. The power of the plasma generator was set at ~100 W. The *h*-BN samples were treated under the same conditions. The scale bars in **(a-d)** represent 20 μm.

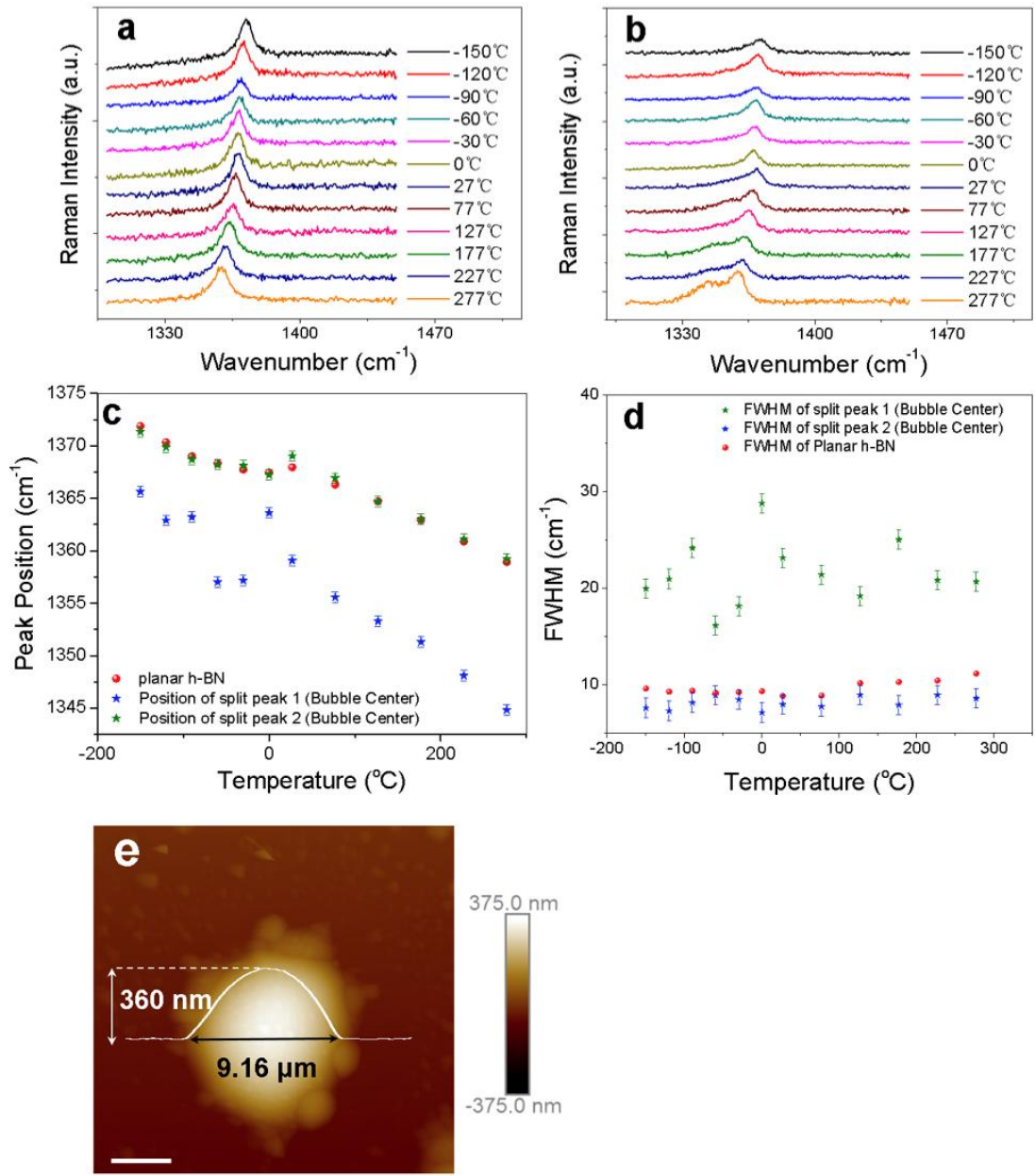

**Supplementary Figure 23 | Temperature dependence of Raman spectra taken of an *h*-BN bubble. a**, Raman spectra taken of the flat surface of an *h*-BN flake at temperatures ranging from -150 ℃ to 277 ℃. **b**, Raman spectra taken of the center of a bubble on an *h*-BN flake in the temperature range from -150 ℃ to 277 ℃. **c**, $E_{2g}$ peak position distribution from the Raman spectra taken of the flat area (red circles) and the bubble center (both blue and green stars, representing two split $E_{2g}$ peak positions) on the *h*-BN surface with respect to the measurement temperature. **d**, FWHM distribution of the Raman spectra taken of the flat area (red circles) and bubble center (both blue and green stars, representing the FWHM of two split $E_{2g}$ peaks) on the *h*-BN surface in relation to the measurement temperature. The error bars indicate the standard deviation. **e**, AFM height image of the bubble examined by Raman spectroscopy, revealing a diameter of ~9.16 μm and a height of ~360 nm. Scale bar, 3 μm.

**Temperature dependence of Raman spectra of *h*-BN bubble**

Raman spectroscopy is a powerful tool for investigating the mechanical and thermal properties of *h*-BN. On the flat area of the *h*-BN samples, only one obvious peak appeared in the Raman spectrum in the range from 1310 cm$^{-1}$ to 1450 cm$^{-1}$, namely, the E$_{2g}$ peak (~1366 cm$^{-1}$). For the *h*-BN bubbles, the *h*-BN E$_{2g}$ peak broadened and exhibited a slight redshift (~1369 cm$^{-1}$). We then gradually increased the annealing temperature of the *h*-BN samples from –150 ℃ to 277 ℃ and recorded Raman spectra of both the flat area and the bubble of *h*-BN at each temperature point. Supplementary Fig. 23a-b shows the typical Raman spectra of *h*-BN in the flat area and the bubble at incremental temperatures from –150 ℃ to 277 ℃. The E$_{2g}$ peak for the flat area of *h*-BN exhibited an obvious redshift, while both a redshift and a split of the E$_{2g}$ peak were observed for the *h*-BN bubble during the heating process. Supplementary Fig. 23c-d show the evolution of the position and FWHM of the peaks.

**Stability study of hydrogen bubbles on *h*-BN**

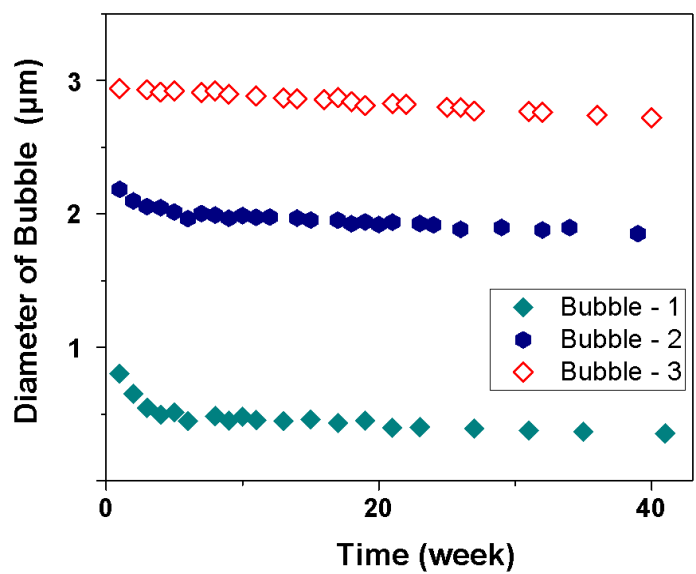

**Supplementary Figure 24 | Stability study of hydrogen bubbles on *h*-BN.** Geometric evolution of hydrogen bubbles with different dimensions under ambient conditions. This experiment is conducted by measuring the diameter-variation of bubbles on *h*-BN (type-851) with an optical microscope.

The bubbles were firstly subjected to a test lasting ~40 weeks to examine the permeability of the pressured *h*-BN bubbles filled with hydrogen molecules. Bubbles with different diameters (~1, 2, and 3 μm) were tested, and the samples were measured by AFM. The geometric evolutions of the hydrogen bubbles with different dimensions are plotted in Supplementary Fig. 24. Some pronounced tendencies of leakage are observed in the beginning period of the test for the bubbles with diameters of ~1 μm and ~2 μm but not for the bubble ~3 μm in diameter, indicating a higher leakage rate for the bubbles with relatively small diameters in the first 8 weeks. The reason for this difference may be that smaller bubbles (~1 and 2 μm) usually have a much higher inner pressure than large bubble (~3 μm) at the very beginning.[12] A slight decrease in the dimension of the large bubble is also observed in the long-term test, with a decrease in the diameter from ~2.94 μm to ~2.72 μm after 40 weeks of hydrogen storage under ambient conditions. Furthermore, the diameters of all the bubbles tended to be stable after ~30 weeks.

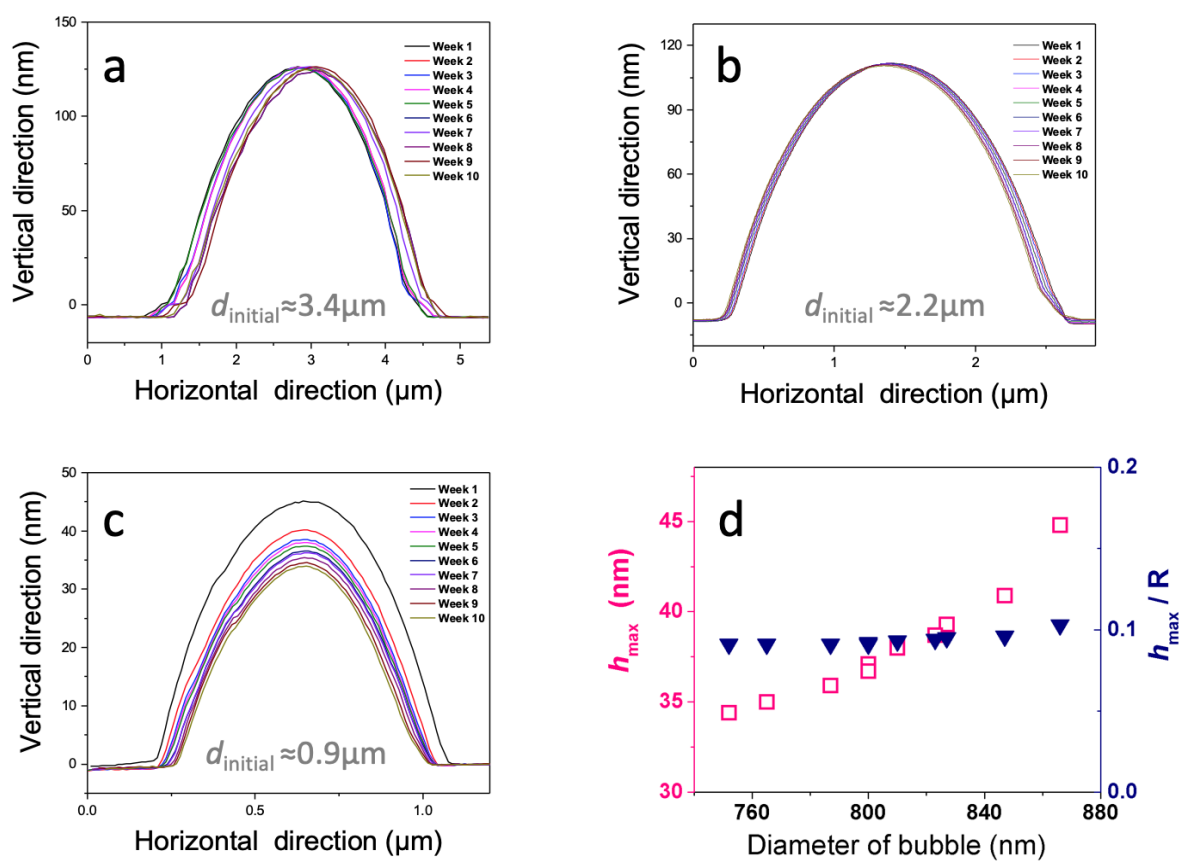

**Supplementary Figure 25 | Stability study of the hydrogen bubbles on *h*-BN by cross-sectional profiles of AFM. a,** A 10-week tracking of the AFM cross-sectional profile of the bubble with a diameter of 3.4 μm. **b,** A 10-week tracking of the AFM cross-sectional profile of the bubble with a diameter of 2.2 μm. **c,** A 10-week tracking of the AFM cross-sectional profile of the bubble with a diameter of 0.9 μm. **d,** Statistics of the aspect ratio ($h_{max}/R$) and the height ($h_{max}$) of the bubble presented in **(c)**

To verify the shrinking-tendency statistics presented in Supplementary Fig. 24, we re-find three new *h*-BN (type-851) bubbles which are similar in dimensions to the former selected bubbles displayed in Supplementary Fig. 24 and re-measure these three bubbles within 10 weeks for about once a week. The results showed a little bit different from our former Supplementary Fig. 25 that both the bubbles with a diameter of 3+ μm and 2+ μm are almost no changed in this 10-week period (Supplementary Fig. 25a and b) while bubbles with a diameter of ~1μm do have some shrinkage in this duration (Supplementary Fig. 25c). We also do the statistics of the aspect ratio ($h_{max}$/R) and the height ($h_{max}$) of the bubble presented in Supplementary Fig. 25c (Supplementary Fig. 25d), showing that the aspect ratio has almost no change in this duration.

The AFM image, Supplementary Fig. 26, showing that a bubble inside a ruptured outer bubble, we realize that gas molecules in *h*-BN bubbles usually have a distribution in the intervals of different layers rather than accumulate in the gap of top 2 layers, which means that larger bubbles are more likely to have a deeper storage of hydrogen than smaller bubbles. We speculate that the latent reason which is responsible for the leakage of small bubbles could be that they are not generated as deep as large bubbles by atomic hydrogen and the gas molecules are accumulated at the superficial intervals (van der Waals gaps) where some defects on the surface are more easily to result in the leakage. However, for some larger bubbles, the gas molecules may distribute in many more gaps of the layered material, in which condition the leakage of some top superficial intervals could be neglected.

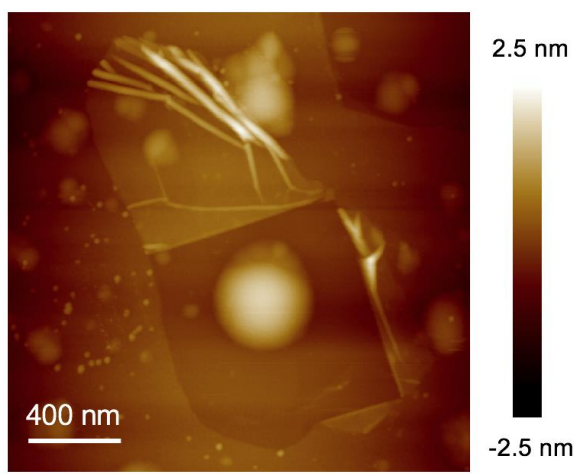

**Supplementary Figure 26 | a bubble inside a ruptured outer bubble.**